\documentclass[referee,extra]{gji} 

\usepackage{timet}
\usepackage{appendix}
\usepackage{graphicx}
\usepackage{subfig}
\usepackage{amsmath}
\usepackage{amssymb} 

\usepackage[dvipsnames,usenames]{color}
\usepackage[colorlinks=true,citecolor=black,linkcolor=black,urlcolor=blue]{hyperref}

\newcommand{\bequ}{\begin{equation} }
\newcommand{\eequ}{\end{equation} }
\newcommand{\eq}{\begin{equation} }
\newcommand{\en}{\end{equation} }

\newcommand{\barr}{\begin{eqnarray} }
\newcommand{\earr}{\end{eqnarray} }
\newcommand{\eqa}{\begin{eqnarray} }
\newcommand{\ena}{\end{eqnarray} }

\newcommand{\bsplitequ}{\begin{split} }
	\newcommand{\esplitequ}{\end{split} }

\newcommand{\be}{ {\bf e } }

\newcommand{\bs}{ {\bf s } }

\newcommand{\by}{ {\bf y } }

\newcommand{\bB}{ {\bf B } }
\newcommand{\bC}{ {\bf C } }

\newcommand{\bH}{ {\bf H } }

\newcommand{\bP}{ {\bf P } }
\newcommand{\bQ}{ {\bf Q } }

\newcommand{\bV}{ {\bf V } }
\newcommand{\bX}{ {\bf X } }

\newcommand{\bZ}{ {\bf Z } }

\newcommand{\bLambda}{ {\bf \Lambda} }

\usepackage{algorithm}
\usepackage{algorithmicx}
\usepackage{algpseudocode} 

\newcommand{\myincludegraphics}[2][]{%
	\includegraphics[#1]{#2}  
}

\begin{document}
	
	\setkeys{Gin}{draft=false}
	
	\title[BFGS-based FWI \& UQ]{Preconditioned BFGS-based Uncertainty Quantification in Elastic Full Waveform Inversion}
	

	\author[Liu et al.]{
		\begin{minipage}{1.\linewidth}
			Qiancheng Liu$^1$, Stephen Beller$^1$, Wenjie Lei$^1$, Daniel Peter$^2$, and Jeroen Tromp$^{1,3}$\\
			{\small~$^1$Department of Geosciences, Princeton University, Princeton, NJ 08544, USA}\\
			{\small~$^2$Division of Physical Sciences and Engineering, King Abdullah University of Science and Technology (KAUST), Thuwal, Saudi Arabia}\\
			{\small~$^3$Program in Applied and Computational Mathematics, Princeton University, Princeton, NJ 08544, USA}
		\end{minipage}
	}
	\date{2020}
	
	\label{firstpage}    
	
	\maketitle

	\begin{summary}
		
		Full Waveform Inversion (FWI) has become an essential technique for mapping geophysical subsurface structures. However, proper uncertainty quantification is often lacking in current applications. In theory, uncertainty quantification is related to the inverse Hessian (or the posterior covariance matrix). Even for common geophysical inverse problems its calculation is beyond the computational and storage capacities of the largest high-performance computing systems. In this study, we amend the Broyden-Fletcher-Goldfarb-Shanno (BFGS) algorithm to perform uncertainty quantification for large-scale applications. For seismic inverse problems, the limited-memory BFGS (L-BFGS) method prevails as the most efficient quasi-Newton method. We aim to augment it further to obtain an approximate inverse Hessian for uncertainty quantification in FWI. To facilitate retrieval of the inverse Hessian, we combine BFGS (essentially a full-history L-BFGS) with randomized singular value decomposition to determine a low-rank approximation of the inverse Hessian. Setting the rank number equal to the number of iterations makes this solution efficient and memory-affordable even for large-scale problems. Furthermore, based on the Gauss-Newton method, we formulate different initial, diagonal Hessian matrices as preconditioners for the inverse scheme and compare their performances in elastic FWI applications. We highlight our approach with the elastic Marmousi benchmark model, demonstrating the applicability of preconditioned BFGS for large-scale FWI and uncertainty quantification. 		
	\end{summary}
	
	
	\section{Introduction}
	
	Seismic full-waveform inversion (FWI) is a compelling approach for characterizing subsurface properties. FWI seeks to estimate the ``optimal'' model by minimizing a measure of data misfit between simulated and observed seismograms, most commonly via an iterative inversion procedure in a least-squares sense~\citep{Lailly1983, Tarantola1984}. Its ultimate goal is to resolve geophysical properties from all available information in observed seismic measurements~\citep{ViOp09,tromp2020nr}. Thanks to advances in data acquisition, high-performance computing, and numerical simulation methods~\citep{Komatitsch1999,peter2011forward,lefebvre201713,polychronopoulou2018broadband}, FWI can constrain seismic models with increasing resolution. Successful applications of FWI across scales have been reported in global~\citep{french2014whole, bozdaug2016global, fichtner2018collaborative, lei2020global}, regional~\citep{tape2010seismic, zhu2012structure, krischer2018automated}, exploration~\citep{warner2016adaptive,metivier2016measuring}, and medical~\citep{bachmann2020source, guasch2020full} imaging. However, due to its ill-posed nature, FWI suffers from non-unique solutions owing to limited data coverage and uncertainties in measurements and theories. Therefore, uncertainty quantification in FWI is essential, but to date only a few solutions have been proposed for larger applications~\citep{Fichtner2011a, Fichtner2011b, Zhu2016, fichtner2018hamiltonian, gebraad2019bayesian, liu2019b, liu2020square, thurin2019ensemble}.
	
	Optimization methods in FWI can be categorized into two families: deterministic and statistical. Deterministic methods, mainly referring to gradient-based optimization~\citep{Pratt1999, Virieux2009}, have been well developed to tackle challenges such as cycle-skipping \citep{warner2016adaptive,metivier2016measuring} and source-encoding~\citep{tromp2019source}. Statistical methods rely on random sampling, based on Markov chain Monte Carlo or Hamiltonian Monte Carlo algorithms~\citep{duane1987hybrid,betancourt2017conceptual}. Such methods are theoretically preferable in FWI, because they provide not only the maximum a posteriori (MAP) model but also statistical metrics for uncertainty quantification~\citep{biswas20172d, fichtner2018hamiltonian, gebraad2019bayesian}. Unfortunately, statistical approaches are often prohibitive for large-scale applications due to their computational expense~\citep{Tarantola2005}. For this reason, deterministic methods are popular in FWI due to their low cost and fast convergence rate. \cite{Tarantola2005} expressed hope for FWI and uncertainty quantification by deterministic optimization if the posterior covariance matrix can be constructed for model appraisal~\citep{tarantola1982generalized}.
	
	Posterior model covariance is closely related to the inverse data-misfit Hessian under the assumption of linearizable forward modeling and Gaussian model priors~\citep{Tarantola2005}. For large-scale applications involving millions of parameters, it becomes unfeasible to store, assemble, and analyze such huge matrices. To tackle this issue, \cite{Zhang1995}~compress the data volume of classic inversion algorithms using least-squares QR factorization. \cite{Trampert2012}~random-probe tomographic models to estimate the resolution length of waveform tomography. \cite{Rawlinson2014}~provide a review of uncertainty estimation in waveform inversion. \cite{Fichtner2015}~analyze direction-dependent resolution lengths from the randomly sampled Hessian via auto-correlation. \cite{Bui-Thanh2013}~approximate the posterior covariance matrix by eigen-decomposing the data-misfit Hessian for its inverse with randomized singular-value decomposition (RSVD)~\citep{Liberty2007,Halko2011}. \cite{Zhu2016}~exploit the point-spread function (PSF) test to improve Hessian-computation efficiency. \cite{Luo2012}~and \cite{liu2019a}~introduced a memory-affordable vector-version square-root variable metric (SRVM) algorithm. Subsequently, \cite{liu2019b}~used SRVM and RSVD to efficiently probe the inverse Hessian for uncertainty quantification in FWI and characterize non-uniqueness based on the SRVM-based null-space shuttle~\citep{thurin2019ensemble,liu2019acoustic,liu2020square}.
	
	In this paper, we explore the feasibility of a classic quasi-Newton method, the Broyden-Fletcher-Goldfarb-Shanno (BFGS) algorithm~\citep{broyden1970convergence,Liu1989}, for elastic FWI with uncertainty quantification. Although the limited-memory BFGS (L-BFGS) algorithm has been used for decades in exploration seismology to invert for subsurface properties, to our knowledge, no attempts have yet been made to reconcile it with uncertainty quantification. Here, we fill this gap by connecting it to the principal BFGS algorithm and combine it with RSVD probing to approximate the posterior covariance matrix of the inverse problem.
	Similar to the SRVM-RSVD workflow~\citep{liu2020square}, we present a BFGS-RSVD approach to access and factorize the inverse Hessian. BFGS runs in the framework of L-BFGS. According to~\cite{Nocedal2006}, L-BFGS is equivalent to the BFGS algorithm if L-BFGS keeps the same initial Hessian and all the memories. Fortunately, FWI usually takes only tens of iterations to converge. The resulting storage of full-memory L-BFGS vectors becomes thus affordable even for large-scale applications.
	
	The performance of BFGS in FWI may be closely related to the initial Hessian guess~\citep{brossier2009seismic,Metivier2013,yang2018time,beller2020probing}. Thus, we investigate the performance of different diagonal approximations of the initial inverse Hessian and compare them in terms of FWI convergence rate and uncertainty quantification maps.
	
	The purpose of this study is to advance L-BFGS and Hessian-related preconditioners for  uncertainty quantification in FWI. We start with a brief review of the FWI optimization problem and recall the theory of the L-BFGS quasi-Newton algorithm. We continue with a presentation of computationally inexpensive diagonal preconditioners used as initial guesses for the L-BFGS approximation of the inverse Hessian. We then discuss retrieval of the inverse Hessian after BFGS-based FWI has converged. Subsequently, we propose a BFGS-RSVD workflow to achieve a faster and cheaper BFGS-based Hessian retrieval. Finally, we verify our method with numerical examples to demonstrate the applicability of preconditioned BFGS-based FWI for uncertainty quantification.
	
	\section{Theory \& Method}
	
	\subsection{BFGS in FWI}
	Seismic FWI aims to iteratively minimize the misfit function~$f\left( \bf{m} \right)$ between observed and synthetic data, ${\bf{d}}$ and~${{\bf{s}}}({\bf{m}})$, respectively. In principle, FWI consists of three consecutive steps: (i) misfit and gradient computations, e.g., with the adjoint-state method~\citep{Tromp2005,Plessix2006}, (ii) a search direction update based on gradients by optimization~\citep{Liu1989,Metivier2013,liu2019a}, and (iii) a linear step search, e.g., using the Wolfe conditions~\citep{wolfe1969convergence,Nocedal2006}, along the search direction. In this section, we focus on the BFGS optimization algorithm, discuss its applicability in FWI, and its potential for uncertainty quantification.
	
	At iteration~$k$, the quasi-Newton search direction~$\bf{p}_k$ is given by
	\eq
	\bf{p}_k =  \mbox{} - \bf{B}_k\nabla f\left( \bf{m}_k \right),
	\en
	where~${{\bf{g}}_k}=\nabla f\left( {\bf{m}_k} \right)$ denotes the gradient, and~${{\bf{B}}_k}$ the inverse Hessian or its approximation. The gradient~${{\bf{g}}_k}$ can be efficiently computed using the adjoint-state method~\citep{Tromp2005,Plessix2006}. For practical applications,  direct computation and storage of the Hessian or its inverse are prohibitive. Instead, L-BFGS provides an efficient matrix-free and iterative approach, which first approximates the Hessian in a rank-two update~\citep{broyden1970convergence,fletcher1970new,goldfarb1970family,shanno1970conditioning}
	\eq
	{{\bf{H}}_{k + 1}} = {{\bf{H}}_k} + \frac{{{{\bf{y}}_k}{\bf{y}}_k^T}}{{{\bf{y}}_k^T{{\bf{s}}_k}}} - \frac{{{{\bf{H}}_k}{{\bf{s}}_k}{\bf{s}}_k^T{\bf{H}}_k^T}}{{{\bf{s}}_k^T{{\bf{H}}_k}{{\bf{s}}_k}}},
	\en
	with~${{\bf{y}}_k} = {{\bf{g}}_{k + 1}} - {{\bf{g}}_k}$ and~${{\bf{s}}_k} = {{\bf{m}}_{k + 1}} - {{\bf{m}}_k}$, and then applies the Sherman-Morrison formula~\citep{sherman1950adjustment} to get the inverse Hessian,
	\eq
	{{\bf{B}}_{k + 1}} = \left( {{\bf{I}} - \frac{{{{\bf{s}}_k}{\bf{y}}_k^T}}{{{\bf{y}}_k^T{{\bf{s}}_k}}}} \right){{\bf{B}}_k}\left( {{\bf{I}} - \frac{{{{\bf{y}}_k}{\bf{s}}_k^T}}{{{\bf{y}}_k^T{{\bf{s}}_k}}}} \right) + \frac{{{{\bf{s}}_k}{\bf{s}}_k^T}}{{{\bf{y}}_k^T{{\bf{s}}_k}}}\,,
	\en
	from which we observe that~${{\bf{B}}_{k + 1}}$ remains in a matrix form, impractical for storage. L-BFGS, which approximates BFGS in a limited amount of memory by considering past gradients and model updates, runs in a two-loop recursion algorithm, shown in Alg.~(1), with~$m$ being the memory value, and~${\gamma}_k$ a scaling factor~\citep{morales2000automatic}.

	\begin{algorithm}[H] 
		\caption{L-BFGS search direction computation}
		\begin{algorithmic}[1]
			\State~$\bf{q} = {\bf{g}}_k$
			\For{$i \gets k-1 \textrm{ to } k-m$}
			\State~$r_i = 1 / {\bf{y}}_i^T{\bf{s}}_i~$
			\State~$a_i = r_i{\bf{s}}_i^T\bf{q}$
			\State~${\bf{q}} = {\bf{q}} - {a_i}{\bf{y}}_i$
			\EndFor
			\State~${\gamma}_k = ({\bf{s}}_{k - 1}^T{\bf{y}}_{k - 1}) /   ({\bf{y}}_{k - 1}^T{\bf{y}}_{k - 1})~$
			\State~${\bf{z}} = \gamma_k {\bf{B}}_k^0{\bf{q}}~$
			\For{$i \gets k-m \textrm{ to } k-1$}
			\State~${b_i} = {r_i}{\bf{y}}_i^T{\bf{z}}$
			\State~${\bf{z}} = {\bf{z}} + {{\bf{s}}_i}({a_i} - {b_i})$
			\EndFor
			\State~${\bf{z}} = \mbox{}-{\bf{z}}$
		\end{algorithmic}
	\end{algorithm}    

	Alg.~(1) outputs the scaled search direction~${\bf{z}} =  \mbox{}-{\gamma _k}{{\bf{B}}_k}{{\bf{g}}_k}$ towards a local minimizer with linear memory requirements. Besides being efficient and inexpensive, this recursion has the advantage that the initial inverse Hessian~${\bf{B}}_k^0$ at iteration~$k$ is included, but isolated from the two-loop recursion. The factor~${\gamma}_k$ attempts to estimate an effective scaling for~$\mbox{}- {{\bf{B}}_k}{{\bf{g}}_k}$, so that a unit step length is accepted for most iterations. ${\bf{B}}_k^0$~opens a window for the preconditioner trials we will conduct for different initial Hessian estimates.
	
	Liu et al.~(2019) and Liu and Peter (2019) discuss the feasibility of Square-Root Variable Metric (SRVM) based FWI and uncertainty quantification. Both SRVM and BFGS belong to the family of quasi-Newton methods. They only differ in that SRVM originates from the Davidon-Fletcher-Powell (DFP) algorithm~\citep{Davidon1959,Fletcher1963}, the dual of BFGS. In an iterative manner, DFP produces a direct approximation to the inverse Hessian while BFGS first approximates the Hessian and then takes its inverse based on the Sherman-Morrison formula. Thus, similar to DFP, we can reconstruct the inverse Hessian from BFGS-based FWI. L-BFGS, a variant of BFGS, has been the state-of-the-art optimization framework for decades in exploration seismology. \cite{Nocedal2006}~state that L-BFGS is equivalent to BFGS if all the memories are kept, while the initial guess of the inverse Hessian remains unchanged. Theoretically, BFGS can capture second-order derivative information from start to finish without dropping histories. As a result, ${\bf{B}}_k^0$ becomes independent of iterations, so we can recast~${\bf{B}}_k^0$ as~${{\bf{B}}_0}$. Thus, there is no need to update the initial Hessian during subsequent iterations. \cite{Modrak2016}~also illustrated that it is unnecessary to regularly update the preconditioner for FWI within a chosen frequency band.
	
	\subsection{Preconditioners in BFGS}
	
	Ideally, the inverse Hessian in elastic FWI can be used to reforge the gradient to directly account for parameter trade-offs as well as source-receiver illumination~\citep{Pratt1999, VirieuxOperto2009, Metivier2013, yang2018time, beller2020probing}. However, explicit computation and storage of the Hessian and its inverse in practical applications are prohibitive. FWI is formulated as a nonlinear minimization of the waveform mismatch between observed and synthetic data via an iterative procedure, indirectly accounting for the inverse Hessian. For large-scale problems~\citep{fichtner2018collaborative, lei2020global}, however, gradients estimated by the adjoint method are also expensive, such that we can only afford a handful of iterations~\citep{tromp2020nr}. As suggested in Alg.~(1), the inverse of an initial Hessian guess can be applied to partially improve performance and reduce the cost of the computationally demanding nonlinear-optimization procedure. Therefore, we first need a direct or indirect estimate of the initial Hessian.
	
	Regarding initial Hessian estimation, in general there are three types of categories: (i) iterative Gauss-Newton, (ii) point-spread function, and (iii) diagonal Hessian estimates. Category (i) involves a Hessian update on the fly~\citep{demanet2012matrix,Metivier2013}; category (ii) is based on point-spread functions for an initial Hessian estimation~\citep{Zhu2016}; category (iii), which is most popular in exploration geophysics, constructs an initial diagonal Hessian using Gauss-Newton methods~\citep{claerbout1971toward,shin2008improved,rickett2003illumination,yang2018time}. We mainly focus on category (iii) due to its effectiveness and ease of implementation, which will be a good starting point for a more sophisticated algorithm, such as L-BFGS, for inverse Hessian construction.
	
	In this study, we take the category~(iii) approach. Given~${\bf{J}} = {{\delta {\bf{s}}} / {\delta {\bf{m}}}}~$ as the first-order Fr\'echet derivative, to approximate the initial inverse Hessian, ${{\bf{B}}_0}$ in Alg.~(1) can be defined as
	\eq
	{{\bf{B}}_0} = {\left[ {\text{diag}\left( {{{\bf{J}}^\dag }{\bf{J}}} \right) + \lambda {\bf{I}}} \right]^{ - 1}} = {\left( {{{\bf{H}}_0} + \lambda {\bf{I}}} \right)^{ - 1}}\,,
	\en
	in which a superscript~$\dagger$ denotes the adjoint, ${{\bf{H}}_0} = \text{diag}\left( {{{\bf{J}}^\dag }{\bf{J}}} \right)$ indicates the diagonal initial Hessian, and the regularization term~$\lambda {\bf{I}}$ exists for stabilization. For convenience, we call preconditioners the initial diagonal approximations of the Hessian although their inverses enter the algorithms.
	
	We can account for the diagonal Hessian in two different ways: simple data illumination~\citep{claerbout1971toward,rickett2003illumination,shin2008improved} and the Gauss-Newton method~\citep{yang2018time}. Each has two alternatives: consideration of only the source geometry or both the source and receiver geometries. Given elastic FWI for P- and S-wave speeds~${\alpha}$ and~${\beta}$, let us start from the well-known source-illumination map~\citep{rickett2003illumination}
	\eq
	{{\rm H}_0(\bf{x})} = \int  \partial_t{\bf{v}}({\bf x},t) \cdot  \partial_t{\bf{v}}({\bf x},t)\,\mathrm{d}t,
	\en
	in which~${\bf{v}}({\bf x},t)$ denotes the source wavefield particle velocity. We impose the same~${{\rm H}_0}({\bf x})$ over the gradients of~${\alpha}$ and~${\beta }$ to obtain our first kind of preconditioner~${\bP_1} = \left\{ {{{\rm H}_0}({\bf x}),{\rm H}_0}({\bf x}) \right\}$. \cite{Luo2012}~derives a similar form called the ``ray density'' preconditioner. Later, we will show in our examples that even this kind of simple preconditioner can lead to a significant improvement in FWI convergence. However, ${H_0}$ fails to account for the acquisition geometry, so we introduce a modified form of eq.~(5), namely,
	\eq
	{\bar {\rm H}_0}{(\bf x)} = \left| \int \partial_t{\bf{v}}({\bf x},t) \cdot \partial_t\bar{\bf{v}}({\bf x},t) \,\mathrm{d}t \right|,
	\en
	in which~${\bf{v}}({\bf x},t)$ and~${\bar {\bf{v}}({\bf x},t)}$ are the source and receiver velocity wavefields, respectively. Following~\cite{Luo2012}, we take the absolute value in eq.~(5) to ensure positive definiteness of the approximate initial Hessian. As a result, our second preconditioner is~${\bP_2} = \left\{ {{{\bar {\rm H}}_0}(\bf{x}),{{\bar {\rm H}}_0}(\bf{x})} \right\}$. The computation of~${\bP_1}$ only involves the source wavefield, whereas that of~${\bP_2}$ involves both the source and receiver wavefields. Therefore, it is expected to see a better performance of~${\bP_2}$ than~${\bP_1}$ in compensation for uneven source-receiver data coverage.
	
	Although~${\bP_1}$ and~${\bP_2}$ can be useful in accelerating FWI, an approximate ``elastic'' initial Hessian should have better performance due to accommodating inherent nonlinearity in elastic inversions. A mathematical derivation for the multi-parameter initial Hessian can be found in Appendix~A. Eq.~(A21) shows that the elastic initial Hessian can be expressed as
	\eq
	\begin{array}{l}
		{{\rm H}_{\alpha \alpha }(\bf{x})} = \int {\left[ {\frac{{\partial {\bf{C}}}}{{\partial \alpha }}\,{\bf{Dv}}({\bf x},t)} \right] \cdot \left[ {\frac{{\partial {\bf{C}}}}{{\partial \alpha }}\,{\bf{Dv}}({\bf x},t)} \right]\,\mathrm{d}t},\\
		{{\rm H}_{\beta \beta }(\bf{x})} = \int {\left[ {\frac{{\partial {\bf{C}}}}{{\partial \beta }}\,{\bf{Dv}}({\bf x},t)} \right] \cdot \left[ {\frac{{\partial {\bf{C}}}}{{\partial \beta }}\,{\bf{Dv}}({\bf x},t)} \right]\,\mathrm{d}t},
	\end{array}
	\en
	where~${\bf{v}}({\bf x},t)$ denotes the particle velocity of the source wavefield, ${\bf{C}}$ the~$\left( {\alpha ,\beta ,\rho } \right)$  related stiffness matrix, and~${\bf{D}}$ a combination of differential operators. Details of~${\bf{C}}$ and~${\bf{D}}$ can be found after eq.~(A7). The resulting preconditioner of the third kind can be expressed as~${{\bf P}_3} = \left\{ {{{\rm H}_{\alpha \alpha }(\bf{x})},{{\rm H}_{\beta \beta }(\bf{x})}} \right\}$. Again, as discussed in eq.~(A22) of Appendix~A, we further extend eq.~(7) to account for the acquisition geometry:
	\eq
	\begin{array}{l}
		{\bar {\rm H}_{\alpha \alpha }(\bf{x})} = \left| {\int {\left[ {\frac{{\partial {\bf{C}}}}{{\partial \alpha }}\,{\bf{Dv}}({\bf x},t)} \right] \cdot \left[ {\frac{{\partial {\bf{C}}}}{{\partial \alpha }}\,{\bf{D}}\bar {\bf{v}}({\bf x},t)} \right]\,\mathrm{d}t} } \right|,\\
		{\bar {\rm H}_{\beta \beta }(\bf{x})} = \left| {\int {\left[ {\frac{{\partial {\bf{C}}}}{{\partial \beta }}\,{\bf{Dv}}({\bf x},t)} \right] \cdot \left[ {\frac{{\partial {\bf{C}}}}{{\partial \beta }}\,{\bf{D}}\bar {\bf{v}}({\bf x},t)} \right]\,\mathrm{d}t} } \right|.
	\end{array}
	\en
	As a result, the preconditioner of the fourth kind becomes~${\bP_4} = \left\{ {{{\bar {\rm H}}_{\alpha \alpha }(\bf{x})},{{\bar {\rm H}}_{\beta \beta }(\bf{x})}} \right\}$. It is expected that~${\bP_4}$  outperforms~${\bP_3}$. As for the computational burden, if we compute the initial Hessian separately, ${\bP_1}$ and~${\bP_3}$ come at the cost of one wavefield simulation, and~${\bP_2}$ and~${\bP_4}$ come at the cost of one gradient computation. Thus, the computations are all cheap compared to the total cost of FWI. When it comes to large-scale applications, we can further minimize the computational cost of the preconditioners by computing them together with the gradient. Also, for each positive-definite preconditioner, a large ratio between its maximum and minimum (similar to a large condition number) can result in numerical instability, which can be alleviated by smoothing and damping~\citep{rickett2003illumination}. Here we use the inverse of the smoothed, damped initial diagonal Hessian~$\text{diag}\left( {{{\bf{J}}^\dag }{\bf{J}}} \right)$ in the form of~${\bP_1}$, ${\bP_2}$, ${\bP_3}$, and~${\bP_4}$, respectively, to take the role of~${{\bf{B}}_0}$ in Alg.~(1).
	
	\subsection{Preconditioned BFGS-based Uncertainty Quantification}
	With the BFGS algorithm and its preconditioners in place, when elastic FWI converges after~$n$ iterations, ${{\bf{B}}_{n + 1}}$ can approximate the inverse Hessian from the past~$n$ histories~\citep{Tarantola2005,Nocedal2006} as
	\eq
	{\bf{H}}^{ - 1} = {\bf{B}}_{n + 1}.
	\en
	From Alg.~(1), we see that the reconstruction of~${{\bf{B}}_{n + 1}}$ involves~${{\bf{B}}_0}$, ${{\bf{s}}_i}$, and~${{\bf{y}}_i}$~$(i=0,1,..,n)$. After preconditioning with~${{\bf{B}}_0}$, the wavefield related pieces of information are mainly embedded in~${\bf{B}}_{n + 1} {\bf{B}}_0^{-1}$ , which starts from the identity~${\bf{I}}$ for stabilization. Note that~${{\bf{B}}_0}$ has been estimated and kept, and~${\bf{B}}_{n + 1} {\bf{B}}_0^{-1}$  rather than~${{\bf{B}}_{n + 1}}$ is retrieved from~${{\bf{s}}_i}$ and~${{\bf{y}}_i}$ .
	
	The retrieval of~${{\bf{B}}_{n + 1}}$ from~${{\bf{B}}_0}$, ${{\bf{s}}_i}$, and~${{\bf{y}}_i}$ can be accomplished with a unit pulse probing vector~$\hat{\be}_j = (0, 0, ..., 0, 1, 0, ..., 0, 0)$ in which the unit pulse ``$1$'' is located at the target column (row) index~$j$ ($j = 1,..,M$, with~$M$ being the model dimension). This is a straight forward procedure, however we will see that it is not a very efficient method. Let us extract the~$j$-th row/column elements from~$\bB_{n+1}$ as depicted in Alg.~(2).
	
	\begin{algorithm}[H] 
		\caption{$ \bB_{n+1}~$ probing}
		\begin{algorithmic}[1]
			\State~${\bf{q}} = {\hat{\be}}_j$
			\For{$i \gets n \textrm{ to } 0$}
			\State~$r_i = 1 / {\bf{y}}_i^T{\bf{s}}_i~$
			\State~$a_i = r_i{\bf{s}}_i^T\bf{q}$
			\State~${\bf{q}} = {\bf{q}} - {a_i}{\bf{y}}_i$
			\EndFor
			\State~${\bf{z}} = {\bf{B}}_0{\bf{q}}~$
			\For{$i \gets 0 \textrm{ to } n$}
			\State~${b_i} = {r_i}{\bf{y}}_i^T{\bf{z}}$
			\State~${\bf{z}} = {\bf{z}} + {{\bf{s}}_i}({a_i} - {b_i})$
			\EndFor
			\State~${\bf{z}} = {\bf{z}}-{\bB_0}{{\hat{\be}}_j}$
		\end{algorithmic}
	\end{algorithm}    
	
	The two-loop algorithm Alg.~(2) outputs the matrix-vector product~${\bf{z}} = \left( {{{\bf{B}}_{n + 1}} - {{\bf{B}}_0}} \right){\hat{\be}}$.  However, it is not flexible for arbitrary element extraction. For example, given a model of size~$M$, we need~$M$ such operations in extracting the~$\bB_{n+1}$ diagonals, which is expensive for uncertainty quantification of large-scale applications. 
	
	Fortunately, randomized SVD~\citep{Liberty2007,Halko2011} provides a more efficient eigendecomposition of large matrices, especially for those with low rank. The method proposed in~\cite{Halko2011} can probe a matrix only with one set of random vectors in a much simpler implementation.	
	Given an elastic FWI with model size~$M$ converges after~$N_r$ iterations, the inverse Hessian~$\bB = {\bf{H}}^{ - 1} = {{\bf{B}}_{n + 1}}$ can span a~$M \times M$ full matrix, which is prohibitive in terms of computation and storage. However, we can probe it with a set of~$M \times N_r$ independent random vectors~$\bX$, similar to the approach of~\cite{liu2019b}, to extract its low-rank form. Single-pass randomized SVD allows for an efficient eigendecomposition of the matrix~$\bB$ of size~$M \times M$, as shown in Alg.~(3).
	
	\begin{algorithm}[H] 
		\caption{Single-pass randomised SVD algorithm}
		\begin{algorithmic}[1]	
			\State~$ \textbf{E} = \textbf{B}  \textbf{X}$ \Comment{Sampling~$\bB$ with~$\bX$ }
			\State~$  \textbf{Q}  \textbf{R} = \textbf{E}$ \Comment{ QR decomposition on~$\textbf{E}$}
			\State~$  {\bf \Omega} (\textbf{Q}^{T}  \textbf{X}) = \textbf{Q}^{T}  \textbf{E}$ \Comment{ Solve for~${\bf \Omega}$ }
			\State~$ \textbf{U} \bLambda \textbf{U}^{T} =  \bf \Omega~$ \Comment{ SVD on~${\bf \Omega}$ }
			\State~$ \textbf{V}  =  \textbf{Q} \textbf{U}~$
			\State~$ \bZ  =  \textbf{V} \bLambda \textbf{V}^T~$
		\end{algorithmic}
	\end{algorithm}    
	
	When setting the target as~${\bf{B}} = {{\bf{H}}^{ - 1}}$, we use~$\bX$ to probe~${{\bf{B}}_{n + 1}} - {{\bf{B}}_0}$, yielding~${\bf{E}} = {{\bf{B}}_{n + 1}}{\bf{X}} - {{\bf{B}}_0}{\bf{X}}$. Here~$\bX$  consists of~$N_r$ independent random vectors, with~$N_r \ll M$. Note that we never write out the full matrix~${\bf{B}}$. We simply need to replace the~$\hat{\be}_j$ in Alg.~(2) with~${\bf{\textit{x}}}_k$, (${\bf{\textit{x}}}_k \in {\bf{X}}$, and~$k = 1,..,N_r$), to have~${\bf{E}}$. When performing the QR decomposition, we only need to keep~$\bQ$ of size~$M \times N_r$. Then, except that~$\textbf{V}$ is of size~$M \times N_r$, all other matrices are of size~$N_r \times N_r$. Finally, Alg.~(3) outputs the inverse Hessian in an SVD form as
	\eq
	\bH^{-1}=\bV \bLambda \bV^T + \bf{B}_0,
	\en
	with~${\bf{V}}$ being an~$M \times N_r$ eigenvector matrix, $\bLambda$ the eigenvalue matrix with~$N_r$ diagonal entries, and ${{\bf{B}}_0}$ the damping term. Eq.~(10) provides a convenient way to access arbitrary elements of the inverse Hessian.
	
	For a linearized inverse problem with Gaussian priors, the relation between the model prior and posterior covariance matrices~$\bC_M$ and~$\bC_m$ can be expressed as~\citep{Tarantola2005,Rawlinson2014}
	\eq
	{{\bf{C}}_M} = {\left( {{{\bf{J}}^\dag }{\bf{C}}_d^{ - 1}{\bf{J}} + {\bf{C}}_m^{ - 1}} \right)^{ - 1}},
	\en
	where {\bf{J}} denotes the Jacobian matrix, and $\bC_d$ the data covariance matrix. ${\bf{C}}_m$ can function as a regularization term in inverse problems \citep{Tarantola2005}. When solving tomographic problems with the least-squares formula, the tuning scalar for balancing the data-misfit and model-misfit contributions can de estimated by the relative scaling between $\bf{C}_m$ and $\bf{C}_d$. Without information gain from the data, we have $\bf{C}_M \rightarrow \bf{C}_m$. Otherwise, for a well-converged inversion, the information gain from the data dominates eq.~(11) as $\bf{C}_M \rightarrow ({{\bf{J}}^\dag }{\bf{C}}_d^{ - 1}{\bf{J}})^{-1}$. Therefore, when seismic tomography converges, following~\cite{Tarantola2005}, \cite{Rawlinson2014}, and \cite{liu2019b}, we choose to express eq.~(11) in a practical way as
	\eq
	{{\bf{C}}_M} = {\bf{C}}_m^{1/2}{\left( {\bf{C}}_m^{1/2}{ {{\bf{J}}^\dag }{\bf{C}}_d^{ - 1}{\bf{J}}}{\bf{C}}_m^{1/2} + \bf{I} \right)^{ - 1}}{\bf{C}}_m^{1/2},
	\en
	In doing so, $\bC_m^{1/2}$ will appear as a transformation matrix along with~${\bf{J}}$ and~${\bf{J}}^\dag$, yielding the approximate inverse Hessian as
	\eq
	{{\bf{H}}^{ - 1}} = {\left( {\bf{C}}_m^{1/2}{ {{\bf{J}}^\dag }{\bf{C}}_d^{ - 1}{\bf{J}}}{\bf{C}}_m^{1/2} + \bf{I} \right)^{ - 1}},
	\en
	which can be obtained using Alg.~(1), noting that ${{\bf{y}}_k}$ changes with $\bf{g}_{k+1}$ and $\bf{g}_k$.  
	Then, based on Alg.~(2), we have
	\eq
	{{\bf{C}}_M} = {\bf{C}}_m^{1/2}({\bf{V\Lambda }}{\bf{V}}^T + {\bf{B}_0}) {\bf{C}}_m^{1/2}.
	\en
	$\bC_m$~and $\bC_M$~denote the prior and posterior model distributions~\citep{Bui-Thanh2013,Zhu2016,liu2019b} and be used to assess the model null-space~\citep{liu2020square}. Note that even without the data information gain from the iterative inversion, ${\bf{B}_0}$ here can serve for a rough uncertainty estimation. The prior $\bC_m$~may be estimated based on geological information and interpretation, well logs, or seismic imaging~\citep{Fomel2003}. Prior information has a significant impact on the inversion. A prior $\bC_m$ that is too general provides little useful information, but one that is too strict may lead to biased inversion results \citep{Tarantola2005}. To simplify the remainder of this paper, we equate~$\bC_m$ to a scaled identity matrix~$\epsilon \bf{I}$, where $\epsilon$ is related to the standard deviation of the model prior. The square-root diagonals of~$\bC_M$, known as the standard deviations, provide a quantitative measure of the posterior distribution~\citep{Tarantola2005}. We can also see from eqs~(12) and (14) that, given a simple~${{\bf{C}}_m}$, the main features of~${{\bf{C}}_M}$ are reflected by~${{\bf{H}}^{ - 1}}$. The approximation of~${{\bf{H}}^{ - 1}}$ via preconditioned-BFGS is the focus of this paper. 
	
	\section{Numerical examples}
	Let us consider the isotropic elastic Marmousi benchmark model, with a parametrization of P- and S-wave speeds and density, as a well-studied application in exploration seismology. We only account for the P- and S- wave speeds~$V_p$ and~$V_S$, but fix the density in the inversion due to its low-sensitivity in FWI~\citep{Virieux2009,blom2017synthetic} without limiting the applicability of our approach. In the following, we aim to show the feasibility of BFGS and preconditioned-BFGS algorithms in elastic FWI, and compare them in terms of convergence rates and uncertainty quantification maps.
	
	\subsection{2D elastic Marmousi benchmark}
	The 2D elastic Marmousi benchmark (Martin et al.~2006) is popular in exploration geophysics due to its substantial structure complexities and wavespeed variations that pose nonlinear challenges in FWI. We use a modified Marmousi here, whose model dimensions are 9,200~m in the horizontal and 3,000~m in the vertical directions, without a water layer. Fig.~1 shows the true and initial elastic models, respectively. Fig.~1b is smoothed from Fig.~1a with a Gaussian blur wide enough to remove discontinuities and distort traveltimes. We run forward and adjoint simulations with a 2D spectral-element code, i.e., SPECFEM2D~\citep{Komatitsch1998,Komatitsch1999}, using absorbing boundaries~\citep{stacey1988improved, Komatitsch2007} around all model edges to mimic wavefield propagation in an infinite domain. The observation system located 10~m underground consists of evenly-distributed 32~shots between 279~m and 8,921~m, and 500~multi-component geophones between 100~m and 9,100~m, respectively. The source time function is a Ricker wavelet of 4~Hz peak frequency. For the simulations, the time step is 0.9~ms, and the recording duration is 7.2~s. As a FWI workflow tool, we use SeisFlows~\citep{Modrak2016}.
	
	We start from running BFGS-based elastic FWI with full-memory L-BFGS. To validate the effectiveness of BFGS in FWI, we compare its inversion results in Fig.~2b with those by L-BFGS with 5 memories in Fig.~2a (a value of~5 is commonly used in L-BFGS approaches). We see that they yield almost identical image results in~$V_P$ and~$V_S$. Fig.~3 further compares the convergence behavior of BFGS and L-BFGS based FWI in terms of normalized data and model misfits. It shows that for our particular example and acquisition setup, BFGS exhibits a slightly faster convergence rate than L-BFGS, but at the price of using the full-history memory of vectors~$\bs_i$ and~$\by_i$ mentioned above. Similar L-BFGS and SRVM comparisons can be found in ~\cite{liu2019a}, where the SRVM algorithm resembles BFGS in keeping all the memory states. Note that~$\bs_i$ and~$\by_i$ sizes are the same as the model size, and their storage increases linearly with iterations. 
	
	Generally, geophysical inverse problems can encounter high nonlinearities, for example a salt-body contrast. In such cases, L-BFGS approaches benefit from lower memory values~\citep{Modrak2016}. However, to alleviate the nonlinearity of the inverse problem, one can resort to approaches using multi-scale inversion strategies~\citep{bunks1995multiscale}, adaptive waveform inversion~\citep{warner2016adaptive}, or optimal-transport metrics~\citep{metivier2016measuring}. Also, it is worth mentioning that a main target of our study is to apply this method in global waveform tomography, which often settles into superlinear convergence rates~\citep{tromp2020nr}.  
	
	The aim of preconditioning in FWI is to accelerate convergence rates. In theory, the optimal preconditioner is the true inverse Hessian itself. Thus, preconditioning aims to find good initial Hessian approximations. Regarding the four preconditioners mentioned above, P1 and P3 only need the forward wavefield while P2 and P4 involve both the source and receiver wavefields. All preconditioners are computed before entering the FWI iterations. Smoothing and damping are required to avoid numerical instabilities. We smooth the Hessian the same way as the gradient, and then stabilize the inverse operation with the median of the smoothed Hessian. Note that one could try to tune the stabilizing coefficient for performance improvements, whereas we fix it based on the minimum and maximum values of the smoothed Hessian. Fig.~4 shows the inversion results of preconditioned BFGS-based FWI with P1, P2, P3, and P4, respectively. Comparison of Figs~2b and 4 indicates that the preconditioned BFGS algorithms perform almost identical, or even better than pure BFGS in driving waveform tomography. The convergence comparisons in Fig.~5 further highlight significant convergence speedups, thanks to the preconditioners. Fig.~5a shows that P3 and P4 outperform P1 and P2, respectively, in reaching similar data misfits but with fewer iterations, as expected. These cross-comparisons stress the importance of accounting for the limited acquisition geometry in preconditioner estimations. 
	
	Besides the stored preconditioners, we keep the set of~$\bs_i$  and~$\by_i$ vectors for the inverse Hessian reconstruction. Even with full-memory histories, $\bs_i$  and~$\by_i$ inflict manageable storage burdens. We will see later that when accessing the approximate inverse Hessian, we only need to fetch small segments of the stored files into memory per operation using a memory-map technique, e.g., with a MemMap algorithm in NumPy, to mitigate the peak memory cost. Afterwards, we can factorize the inverse Hessian by running randomized SVD over the stored preconditioner and the~$\bs_i$  and~$\by_i$ vectors. During this process, we never write out the full matrix of the inverse Hessian thanks to randomized SVD. As discussed previously and in~\cite{liu2019b}, the rank of the approximate inverse Hessian equals the number of iterations~$N_{iter}$. The resulting benefit is that we only need~$N_{iter}$ such independent random vectors of model size for the random probing~$E = ZX$ to start Alg.~(3), which subsequently outputs the inverse Hessian in SVD form. The eigen-order number of the factorized matrix is then the same as the iteration number.
	
	The top of Figs.~6 to~10 shows the eigen-spectrum of the inverse Hessians from the BFGS-based elastic FWIs. Each of them starts from values with similar magnitudes, but ends differently depending on the level of convergence. Their corresponding eigenvector groups are displayed following each eigen-spectrum. Each group has five eigenvectors of increasing eigen-orders, including the 1st, 5th, 10th, 20th, and final one. All eigenvectors per group are orthogonal. In each eigenvector group, we see that as the eigen-order increases, the energies begin to move up towards the observation system. We derive this tendency of the energy moving from the inverse of the inverse Hessian~$\bB$, i.e., the Hessian~$\bH$, whose eigen-energies fade away from the observation system as the eigen-order increases~\citep{Bui-Thanh2013,Zhu2016}. Another intriguing observation is that there is some random noise in Figs.~7f to~10f. When looking at the corresponding eigen-spectra, we interpret this as due to rank deficiency. It means that the actual rank of this matrix is less than the pre-estimated number, which has the advantage that one can safely truncate the SVD to further save storage.
	
	With the inverse Hessians in SVD form, we can efficiently extract their standard deviations for uncertainty quantification maps, as shown in Fig.~11. Overall, they all look similar and the features of the uncertainty maps resemble those by SRVM-based FWI~\citep{liu2019b} and Ensemble Kalman Filtering (EnKF)~\citep{thurin2019ensemble}. Noticeable is that the methods reported here are based on the most popular quasi-Newton method, L-BFGS, without much additional cost. To understand the uncertainty maps from a physical perspective, again, we can consider the data coverage, which directly reflects the Hessian. Data coverage decreases due to geometrical spreading and, specifically, a high-wavespeed anomaly may deflect energy. We can thus infer that the appearance of the uncertainty maps should counteract the Hessian characteristics. The uncertainty maps in Fig.~11 reflect multiple features of the inverse Hessian, in particular: (1) uncertainty increases as data coverage decreases; (2) high-wavespeed structures are often related to relatively high uncertainties, especially for the two salts in the corners. In short, the uncertainty map indicates that the model space with better data coverage will have more information gain, i.e., smaller uncertainties.
	
	
	For a detailed investigation of the uncertainty maps, we first look at Fig.~11 horizontally to compare the~$V_p$ row with the~$V_s$ row, seeing that the~$V_p$ model usually has larger uncertainties than~$V_s$. We explain this from the perspective that in elastic isotropic media the radiation pattern of~$V_p$ is isotropic while that of~$V_s$ is far-offset dominant. FWI favors far-offset data because it can produce gradients of lower-wavenumber than near-offset data. Such lower-wavenumber components, usually with stronger amplitudes, can further help FWI to overcome local minima, to some extent. The vertical row comparison in Fig.~11 shows uncertainty maps from BFGS-based FWIs with different preconditioners. We note that all preconditioners in BFGS result in almost identical~$V_p$ and~$V_s$ maps, however at different convergence rates. We conclude that preconditioners estimated from the linear Gauss-Newton method outperform those purely derived from data illumination. Furthermore, the preconditioners which consider the recording geometry outperform those which do not. Finally, we regard the salt bodies as good identifiers to assess the quality of the presented uncertainty maps, because for those areas we expect to see higher uncertainty values. Based on the above considerations, we infer that among the presented preconditioners the P4 preconditioner has the most advantages, providing reasonable uncertainty maps in Fig.~11e, and the fastest convergence rates in Fig.~5.
	
	Uncertainty quantification provides an a posteriori assessment of the model. To further validate the uncertainty maps in Fig.~11, we show in Fig.~12 the absolute model difference between the true model in Fig.~1 and one inverted model in Fig.~4d. We observe that Fig.~12 resembles Fig.~11, especially Figs.~11c and 11e with preconditioners P2 and P4.   
	
	Overall, we see that preconditioned BFGS algorithms demonstrate considerable speedup in FWI and yield reasonable accompanying uncertainty maps. It is worth mentioning that our preconditioned BFGS algorithms can readily be incorporated into standard FWI workflows. The extra costs in computation and storage before, during, and after waveform inversion are all manageable even for large-scale inversions. This makes our method suitable for exascale geophysical applications, addressing uncertainty quantification in, e.g., global-scale waveform inversions~\citep{bozdaug2016global, fichtner2018collaborative, lei2020global}.
	
	\section{Discussion and Conclusions}
	Uncertainties inherently exist in geophysical inverse problems, such as waveform tomography, due to limitations in observations, theories, and algorithms. The L-BFGS algorithm has become the most popular optimization method in applied mathematics, including for seismic waveform tomography. We have demonstrated that the BFGS algorithm has the potential to address uncertainty quantification in FWI. Furthermore, well-specified preconditioners can help gain significant computational savings, yielding the same, or even better, inversion results. The estimation of uncertainty maps only requires the storage of the related preconditioner and a set of memory vectors in L-BFGS. Based on the variable-metric component of the BFGS algorithm during the inversion process, these stored vectors inherently communicate multi-parameter information from the initial model to the inverted one.
	
	Note that we consider the research here as a linearized inverse problem. Linearization means that we linearize the forward modeling operator around the maximum a posterior (MAP) point. To guarantee the linearization some essential data processing such as data quality control and data selection may be required for the real data. When coming across high nonlinearities, one can resort to the multi-scale inversion strategies and so on for alleviation. Because the uncertainty map needs a set of L-BFGS memory vectors, we hope the inversion can converge smoothly for a given objective function.
	
	The combination of BFGS and randomized SVD methods facilitates the retrieval of a low-rank representation of the inverse Hessian, rather than the Hessian \citep{Bui-Thanh2013,Zhu2016}, from which standard deviations provide a straightforward assessment of inversion convergence and uncertainty. Our approach is strictly based on a standard FWI workflow, augmenting it to a Bayesian inversion under the assumption of linearized forward modeling and Gaussian model priors. Finally, the presented BFGS-based FWI and uncertainty quantification workflows are fully scalable and readily amenable to seismic exascale applications.

	\section{Acknowledgements}
	The authors are grateful to editor Carl Tape and reviewer Christian Boehm and an anonymous reviewer for improving the initial manuscript. The authors are grateful to Frederik J. Simons for inspiring discussions. This research used resources of the Oak Ridge Leadership Computing Facility, which is a DOE Office of Science User Facility supported under Contract DE-AC05-00OR22725.
	
	
	\bibliography{biblio}
	
	%
	%
	%
	\clearpage
	
	\begin{figure}
		\begin{center}
			\captionsetup[subfloat]{farskip=2pt,captionskip=2pt}
			\subfloat[]{%
				\myincludegraphics[width=0.49\textwidth]{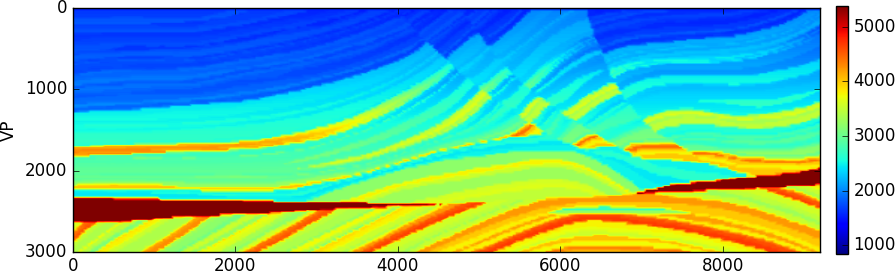}
				\myincludegraphics[width=0.49\textwidth]{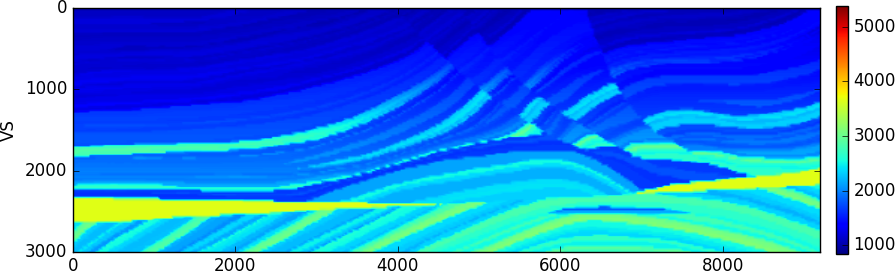}
			}\\
			\subfloat[]{%
				\myincludegraphics[width=0.49\textwidth]{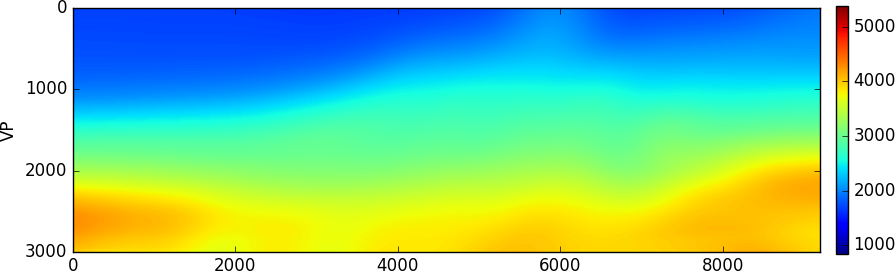}
				\myincludegraphics[width=0.49\textwidth]{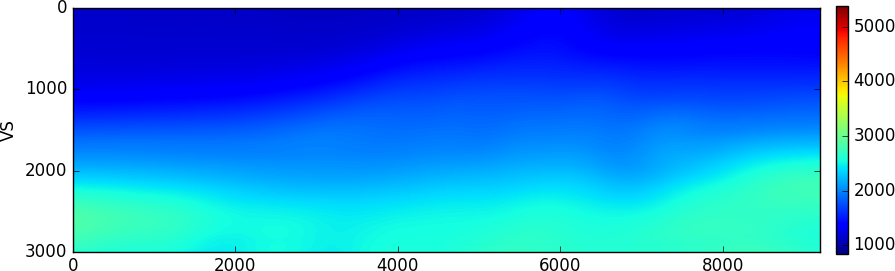}
			}
		\end{center}
		
		\caption{(a) True and (b) smoothed, initial elastic Marmousi models used in elastic FWI. (Left column:~$V_P$\,; right column:~$V_S$\,.)}
	\end{figure}
	
	\begin{figure}
		\begin{center}
			\captionsetup[subfloat]{farskip=2pt,captionskip=2pt}
			\subfloat[]{%
				\myincludegraphics[width=0.49\textwidth]{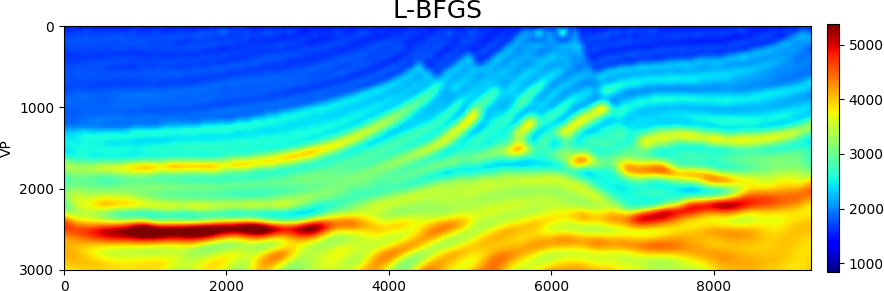}
				\myincludegraphics[width=0.49\textwidth]{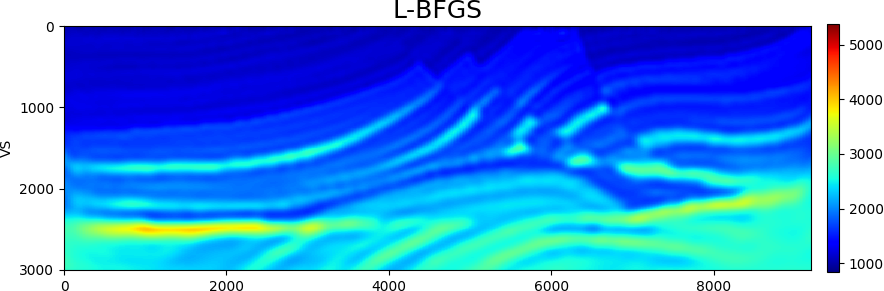}
			}\\
			\subfloat[]{%
				\myincludegraphics[width=0.49\textwidth]{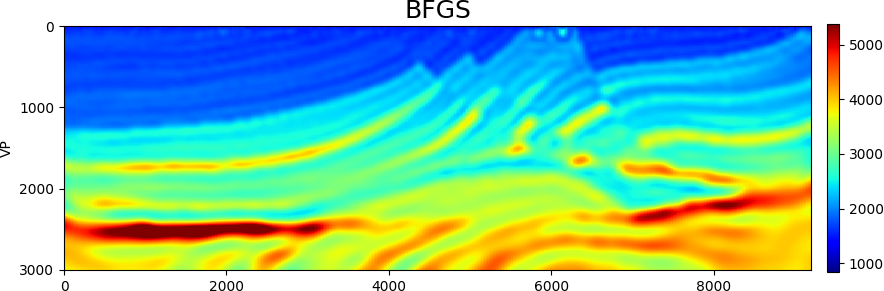}
				\myincludegraphics[width=0.49\textwidth]{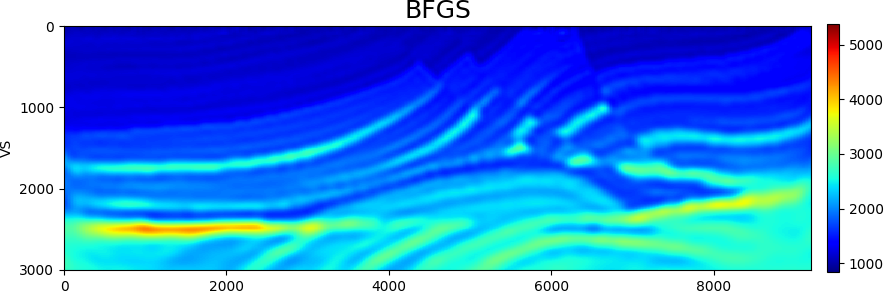}
			}
		\end{center}
		
		\caption{Inverted models from elastic FWI using (a) L-BFGS and (b) BFGS after~$82$ and~$65$ iterations. L-BFGS runs with 5 memories; BFGS runs in the L-BFGS flow but with all the memories. The results look similar even with different iterations. (Left column:~$V_P$\,; right column:~$V_S$\,.)}
	\end{figure}
	
	\begin{figure}
		\begin{center}
			\captionsetup[subfloat]{farskip=2pt,captionskip=2pt}
			\subfloat[]{%
				\myincludegraphics[width=0.45\textwidth]{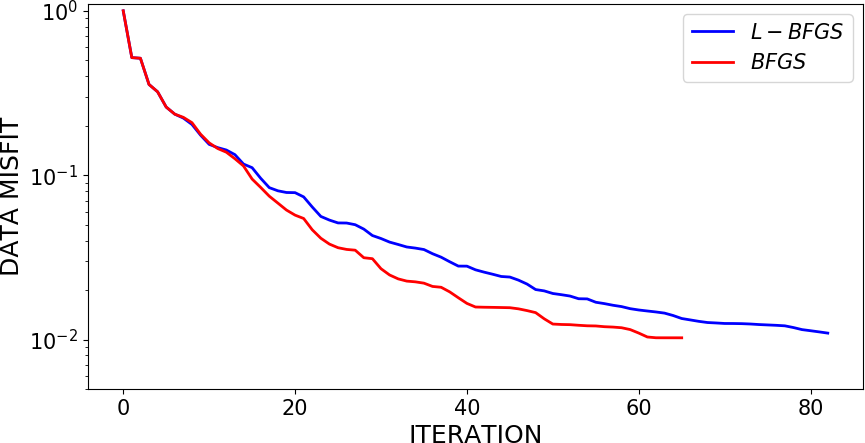}
			}\\
			\subfloat[]{%
				\myincludegraphics[width=0.45\textwidth]{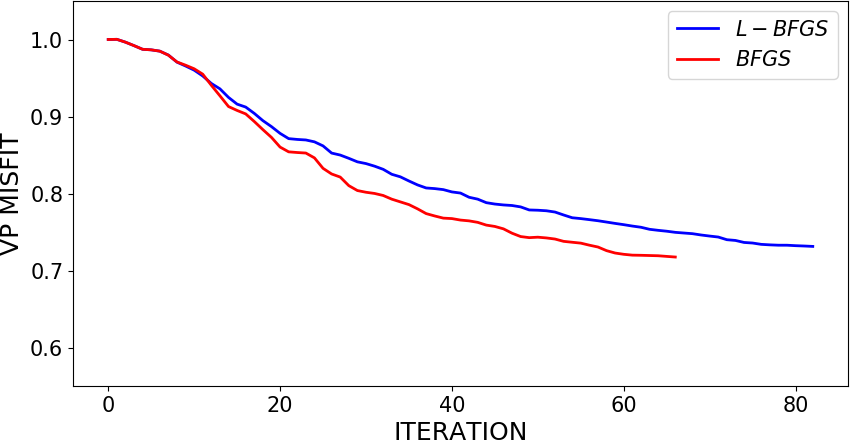}\hspace*{5mm}
				\myincludegraphics[width=0.45\textwidth]{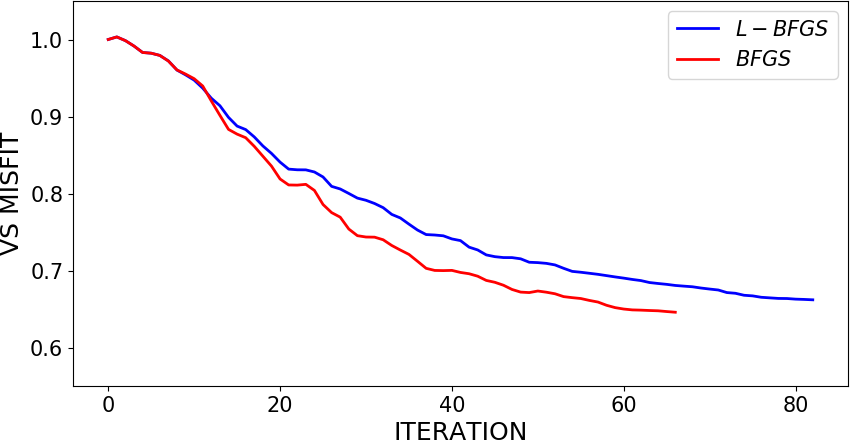}
			}
		\end{center}
		
		\caption{Convergence rates of L-BFGS- and BFGS-based elastic FWI. Plotted are (a) data-misfit and (b) model-misfit comparisons.}
	\end{figure}
	
	\begin{figure}
		\begin{center}
			\captionsetup[subfloat]{farskip=2pt,captionskip=2pt}
			\subfloat[]{%
				\myincludegraphics[width=0.49\textwidth]{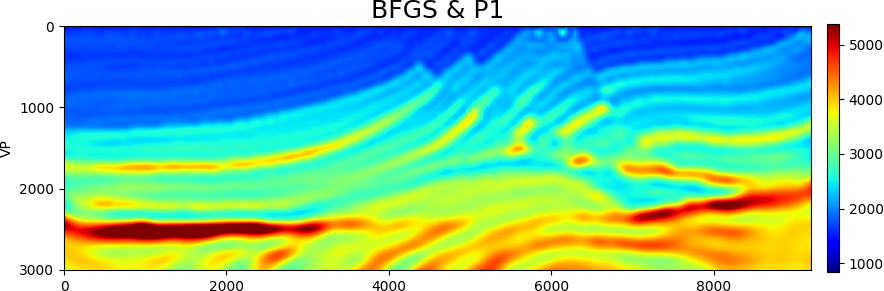}
				\myincludegraphics[width=0.49\textwidth]{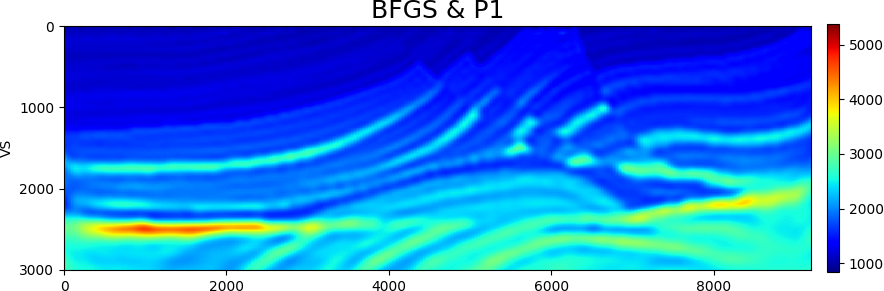}
			}\\
			\subfloat[]{%
				\myincludegraphics[width=0.49\textwidth]{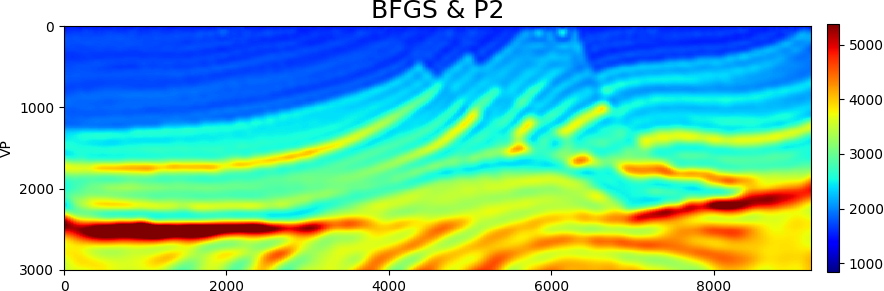}
				\myincludegraphics[width=0.49\textwidth]{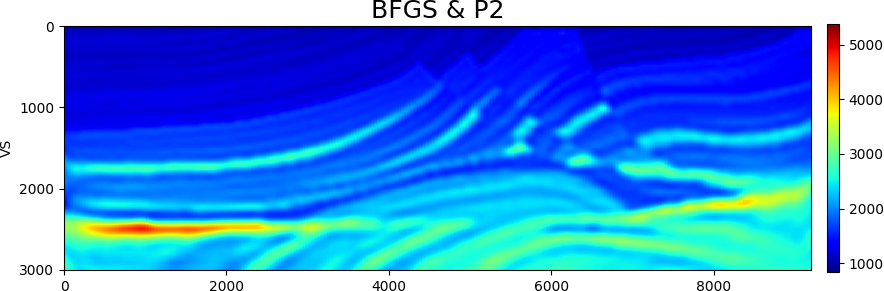}
			}\\
			\subfloat[]{%
				\myincludegraphics[width=0.49\textwidth]{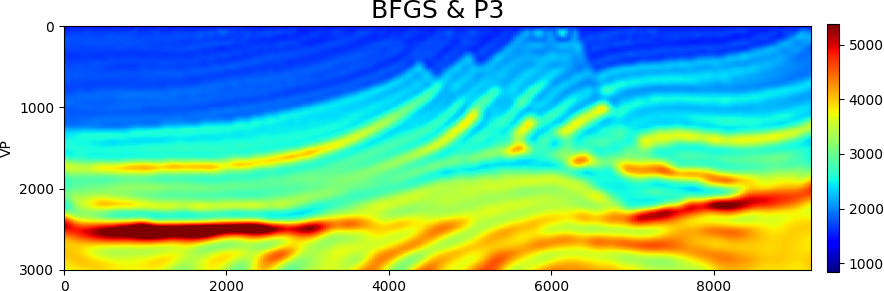}
				\myincludegraphics[width=0.49\textwidth]{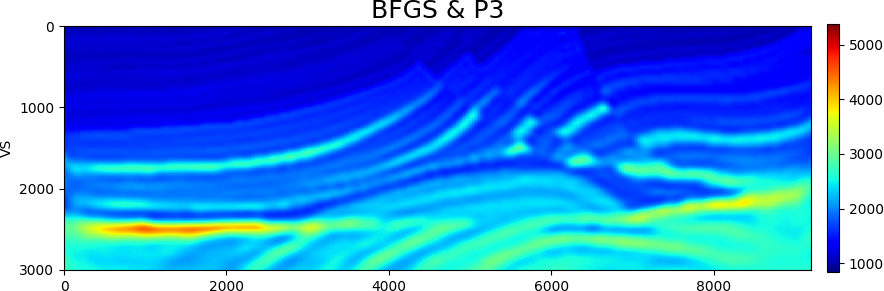}
			}\\
			\subfloat[]{%
				\myincludegraphics[width=0.49\textwidth]{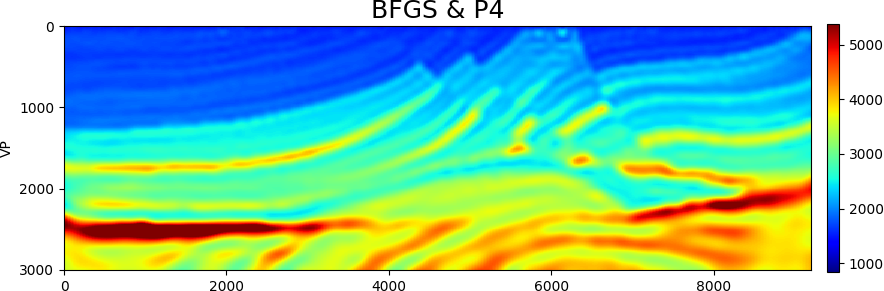}
				\myincludegraphics[width=0.49\textwidth]{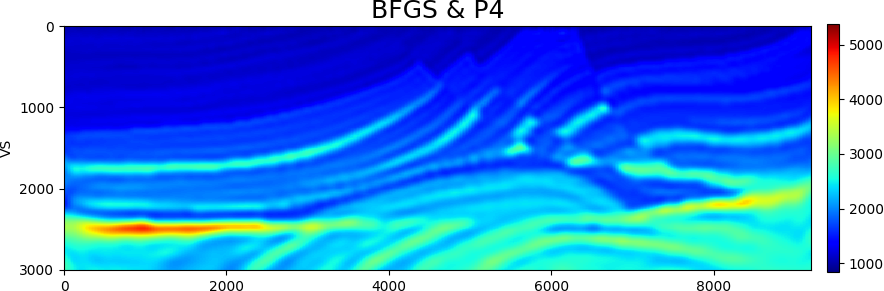}
			}
		\end{center}
		
		\caption{From top to bottom: inverted results of BFGS-based FWI with four different preconditioners after~$81$, $81$, $67$, $80$ iterations. Among them, (d) is closest to the true model. (Left column:~$V_P$\,; right column:~$V_S$\,.)}
	\end{figure}
	
	\begin{figure}
		\begin{center}
			\captionsetup[subfloat]{farskip=2pt,captionskip=2pt}
			\subfloat[]{%
				\myincludegraphics[width=0.45\textwidth]{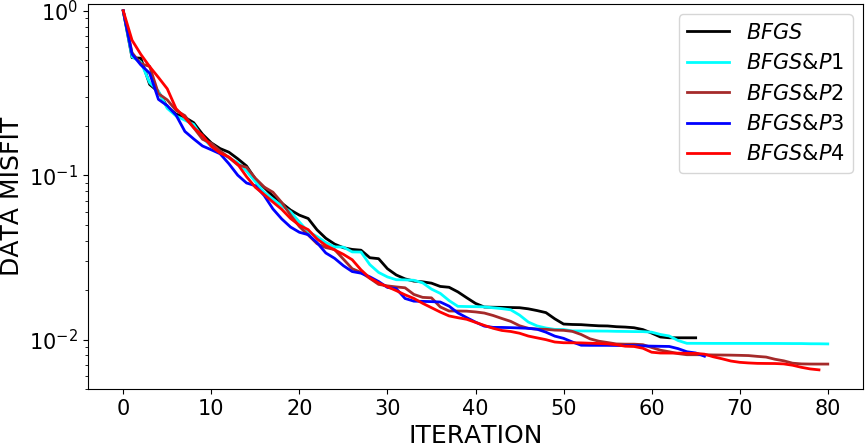}
			}\\
			\subfloat[]{%
				\myincludegraphics[width=0.45\textwidth]{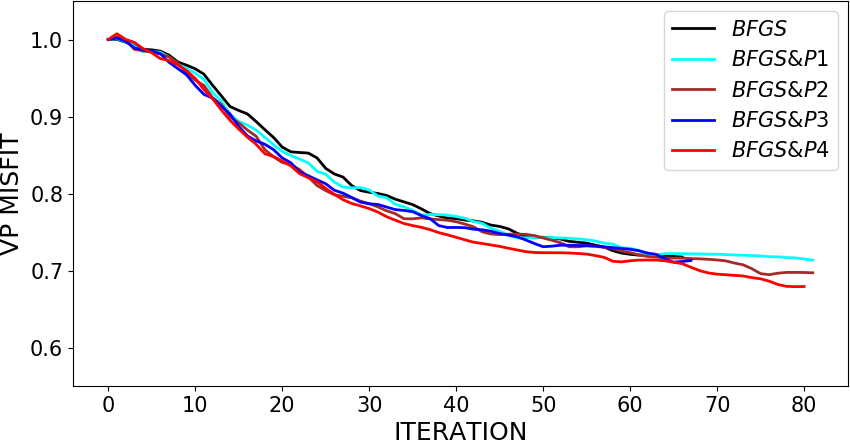}\hspace*{5mm}
				\myincludegraphics[width=0.45\textwidth]{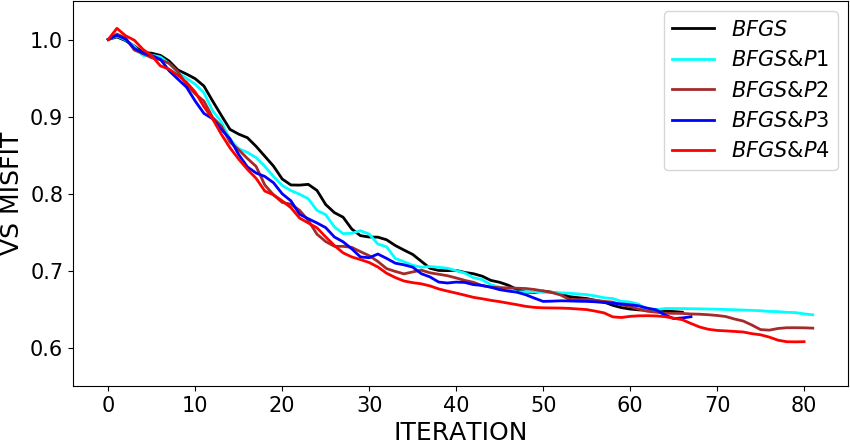}
			}
		\end{center}
		
		\caption{Convergence rate comparisons of preconditioned BFGS-based FWI, with the pure BFGS-based inversion in black for reference. Plotted in (a) and (b) are data-misfit and model-misfit convergence curves, respectively.}
	\end{figure}
	
	\begin{figure}
		\begin{center}
			\captionsetup[subfloat]{farskip=2pt,captionskip=2pt}
			\subfloat[]{%
				\myincludegraphics[width=0.42\textwidth]{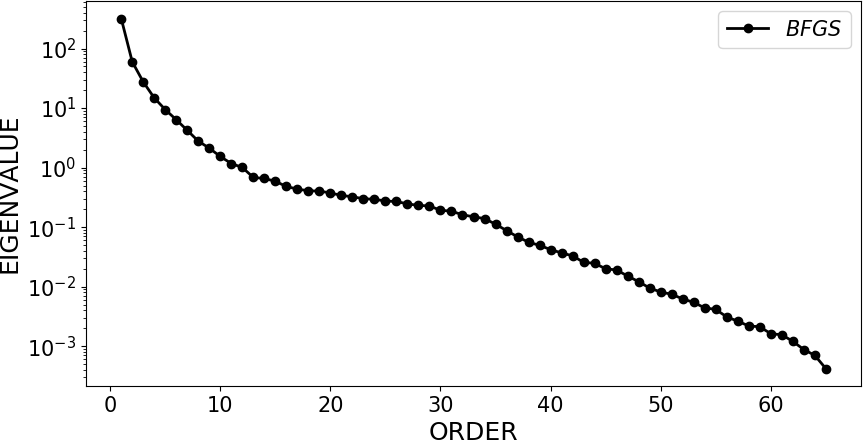}
			}\\
			\subfloat[]{%
				\myincludegraphics[width=0.49\textwidth]{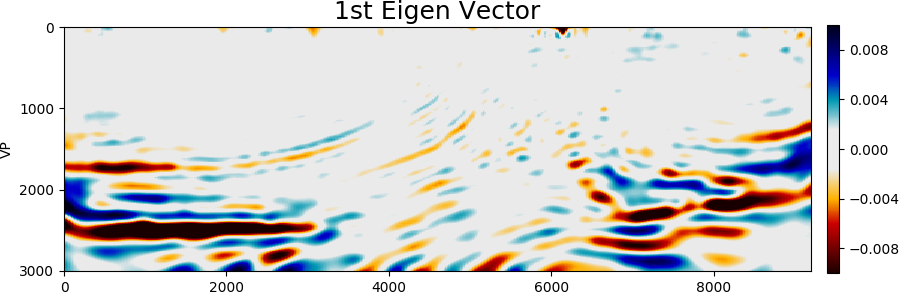}
				\myincludegraphics[width=0.49\textwidth]{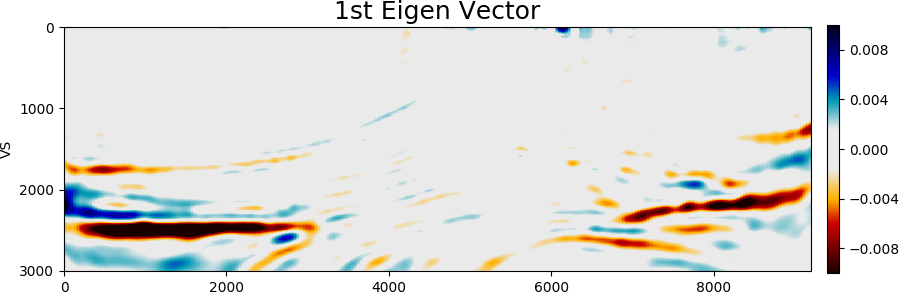}
			}\\
			\subfloat[]{%
				\myincludegraphics[width=0.49\textwidth]{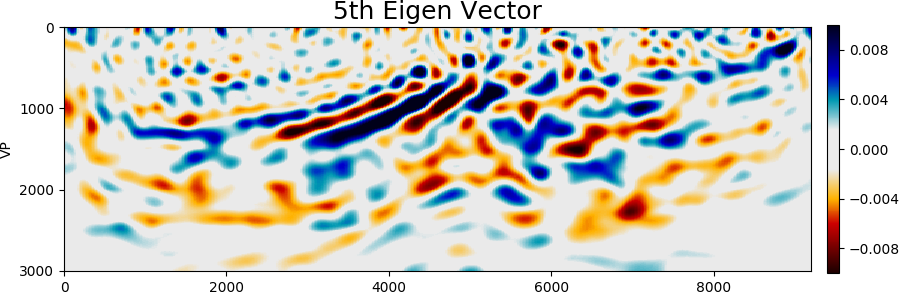}
				\myincludegraphics[width=0.49\textwidth]{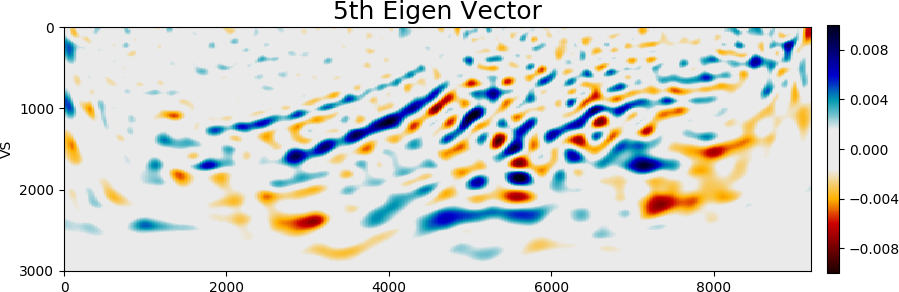}
			}\\
			\subfloat[]{%
				\myincludegraphics[width=0.49\textwidth]{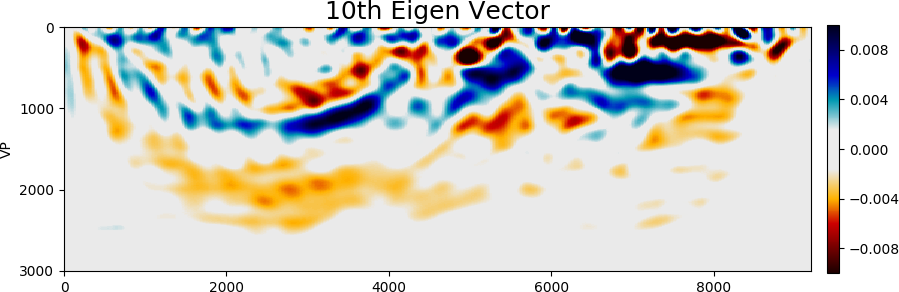}
				\myincludegraphics[width=0.49\textwidth]{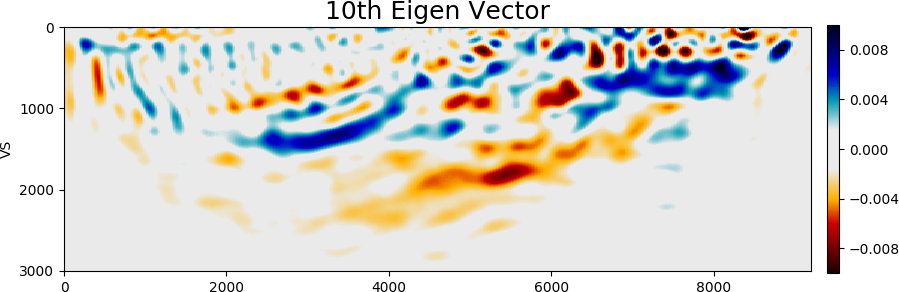}
			}\\
			\subfloat[]{%
				\myincludegraphics[width=0.49\textwidth]{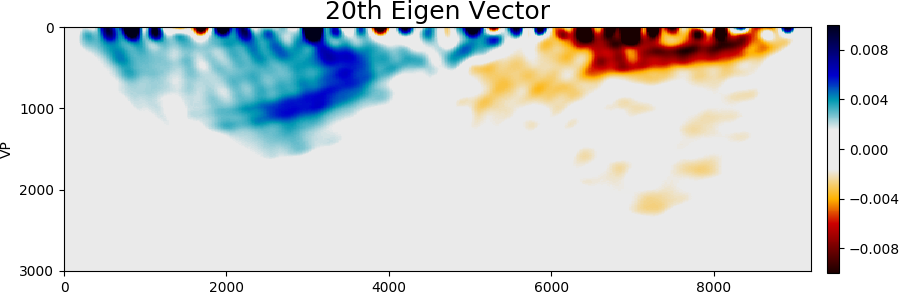}
				\myincludegraphics[width=0.49\textwidth]{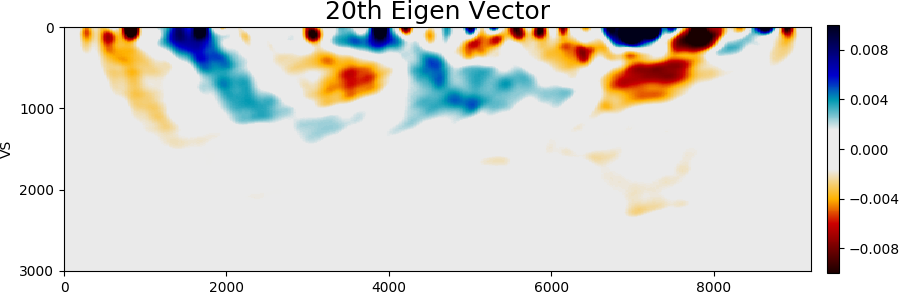}
			}\\
			\subfloat[]{%
				\myincludegraphics[width=0.49\textwidth]{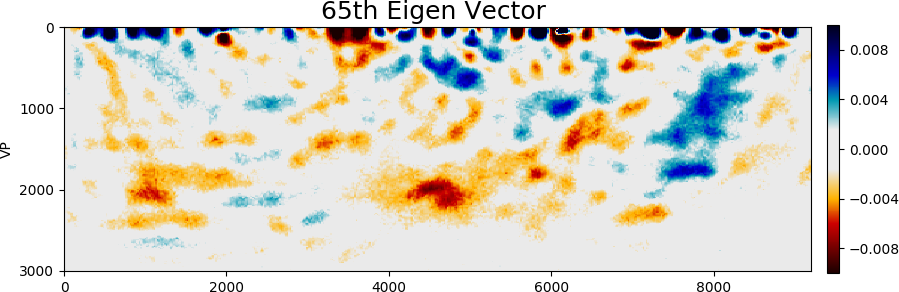}
				\myincludegraphics[width=0.49\textwidth]{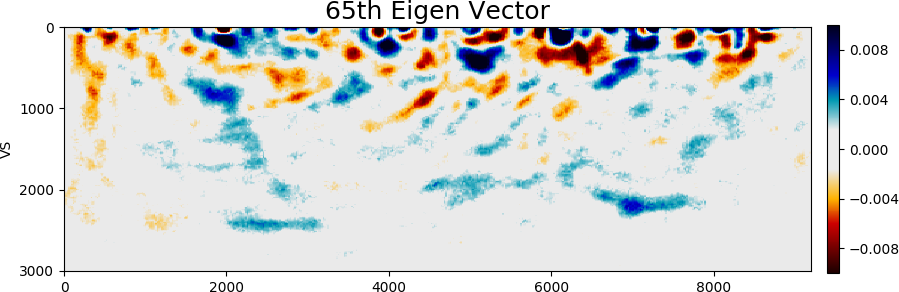}
			}
		\end{center}
		
		\caption{Eigendecomposition of the inverse Hessian from BFGS-based elastic FWI using randomized SVD. Plotted in (a) are the eigenvalues on a logarithmic scale. Plotted in the remainder are the 1st, 5th, 10th, 20th, and the final eigenvectors, respectively. (Left column:~$V_P$\,; right column:~$V_S$\,.)}
	\end{figure}
	
	\begin{figure}
		\begin{center}
			\captionsetup[subfloat]{farskip=2pt,captionskip=2pt}
			\subfloat[]{%
				\myincludegraphics[width=0.42\textwidth]{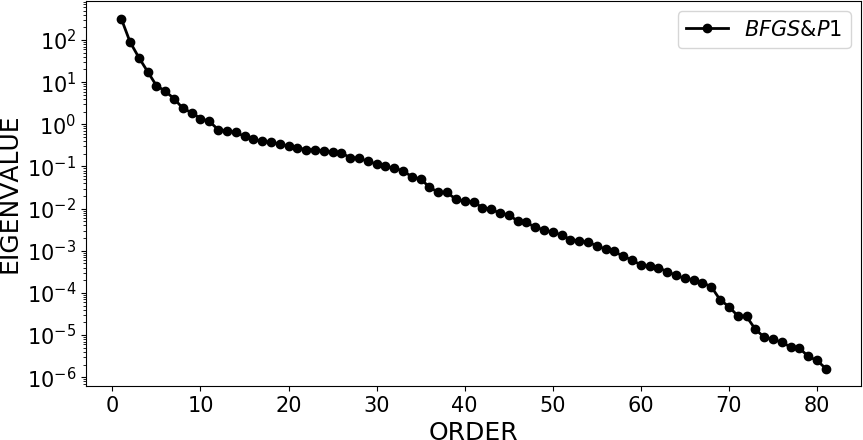}
			}\\
			\subfloat[]{%
				\myincludegraphics[width=0.49\textwidth]{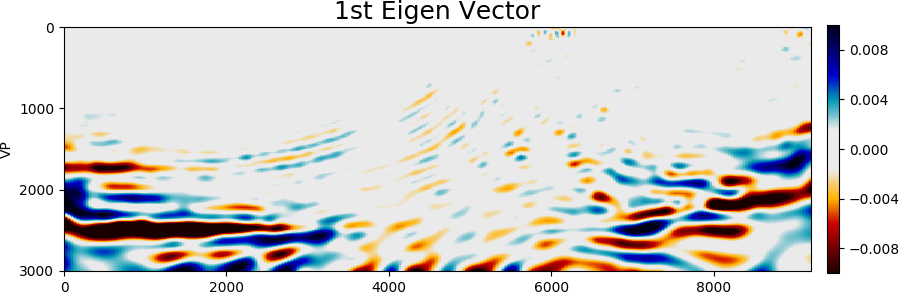}
				\myincludegraphics[width=0.49\textwidth]{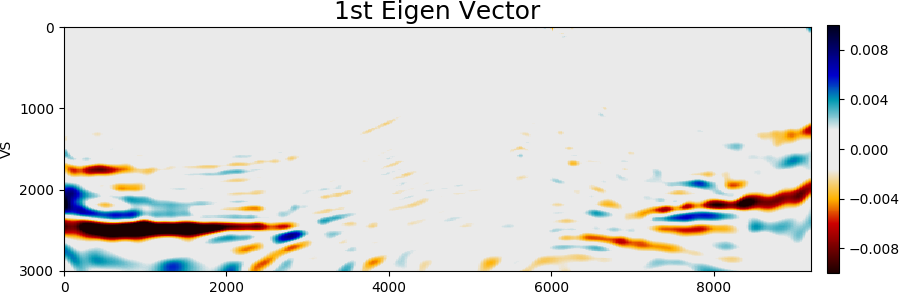}
			}\\
			\subfloat[]{%
				\myincludegraphics[width=0.49\textwidth]{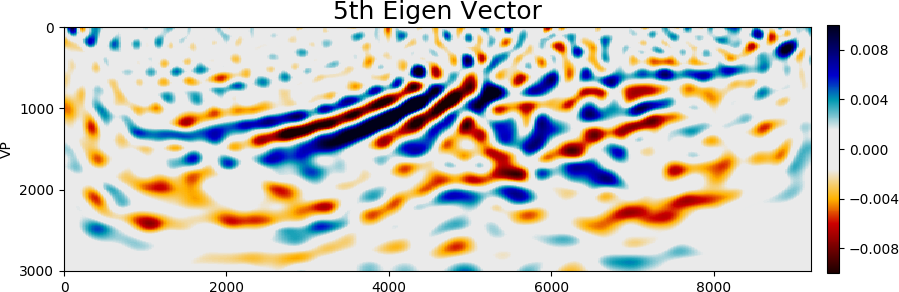}
				\myincludegraphics[width=0.49\textwidth]{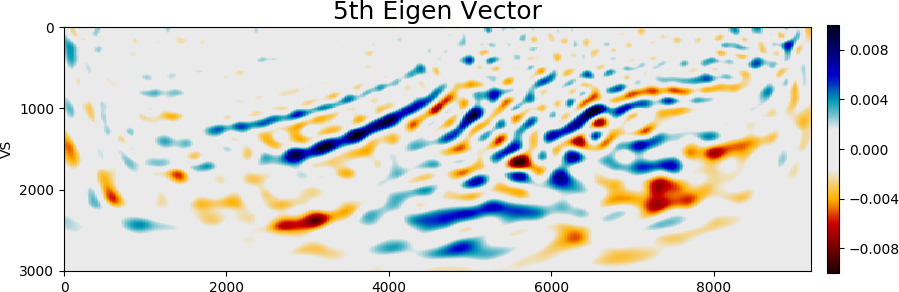}
			}\\
			\subfloat[]{%
				\myincludegraphics[width=0.49\textwidth]{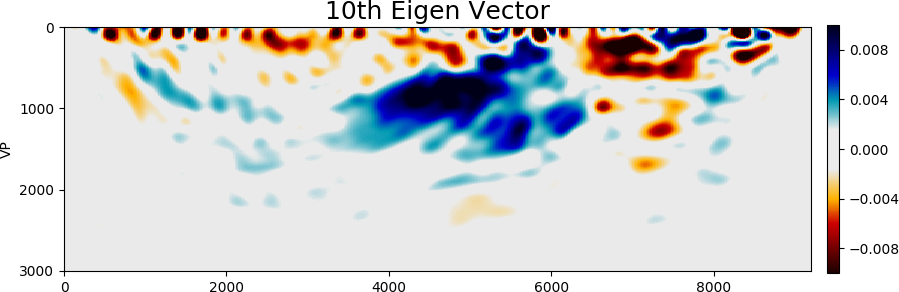}
				\myincludegraphics[width=0.49\textwidth]{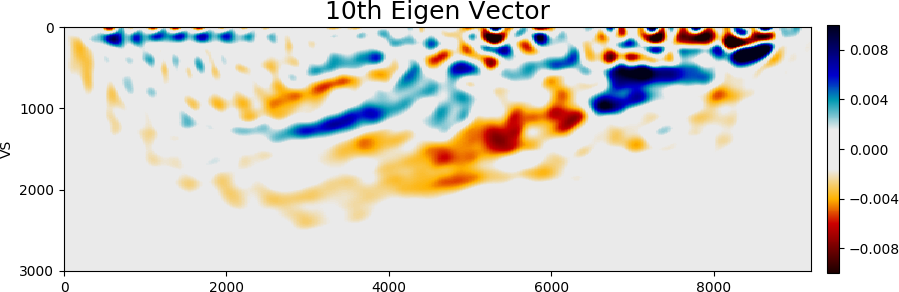}
			}\\
			\subfloat[]{%
				\myincludegraphics[width=0.49\textwidth]{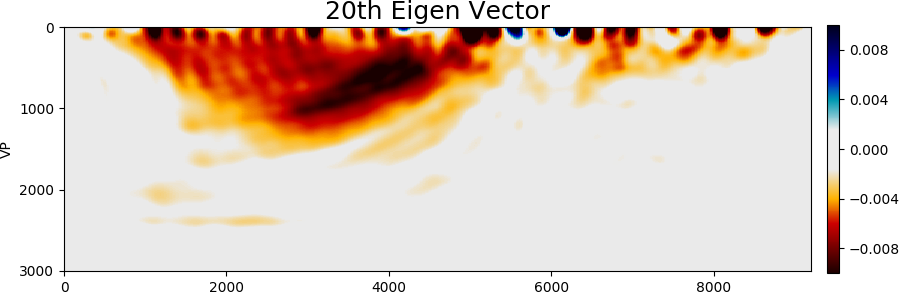}
				\myincludegraphics[width=0.49\textwidth]{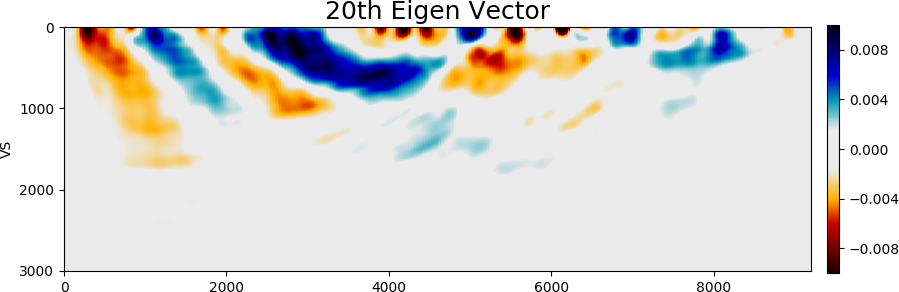}
			}\\
			\subfloat[]{%
				\myincludegraphics[width=0.49\textwidth]{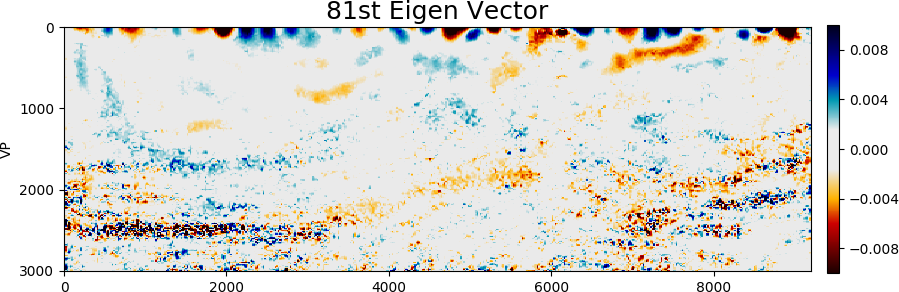}
				\myincludegraphics[width=0.49\textwidth]{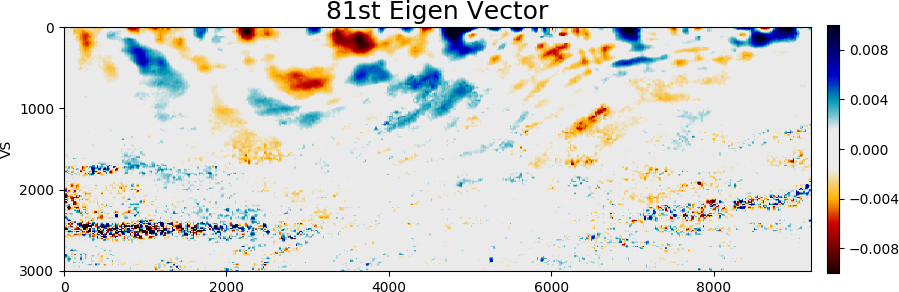}
			}
		\end{center}
		
		\caption{Eigendecomposition of the inverse Hessian from BFGS~\&~P1 elastic FWI using randomized SVD. Plotted in (a) are the eigenvalues on a logarithmic scale. Plotted in the remainder are the 1st, 5th, 10th, 20th, and the final eigenvectors, respectively. (Left column:~$V_P$\,; right column:~$V_S$\,.)}
	\end{figure}
	
	\begin{figure}
		\begin{center}
			\captionsetup[subfloat]{farskip=2pt,captionskip=2pt}
			\subfloat[]{%
				\myincludegraphics[width=0.42\textwidth]{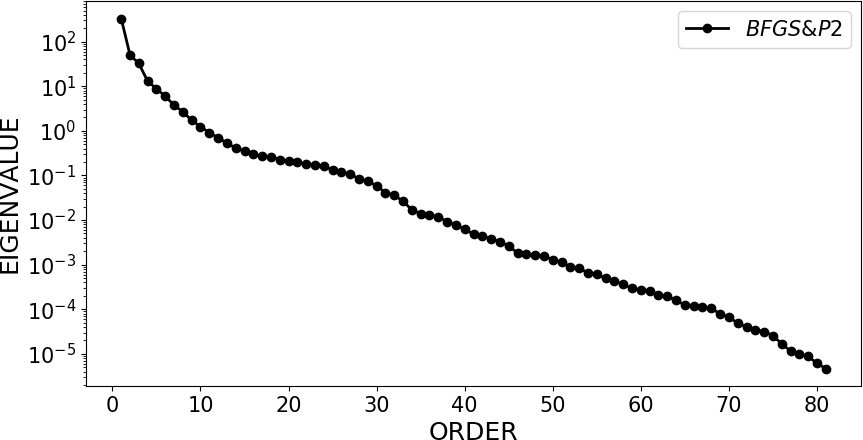}
			}\\
			\subfloat[]{%
				\myincludegraphics[width=0.49\textwidth]{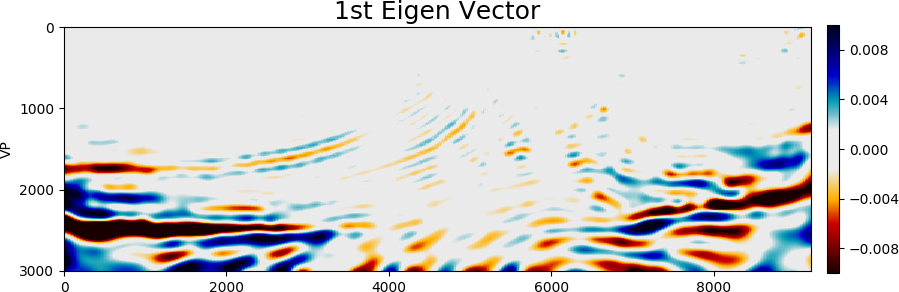}
				\myincludegraphics[width=0.49\textwidth]{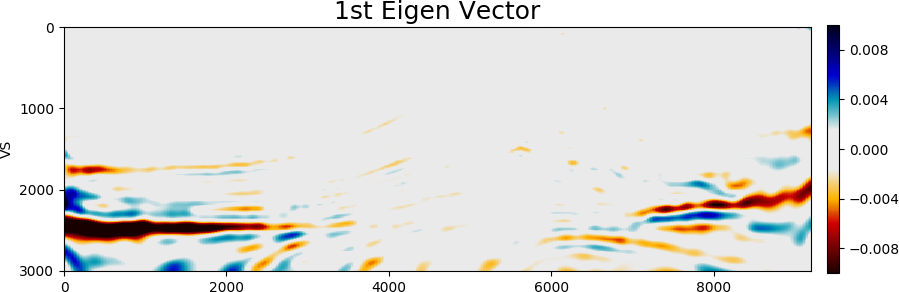}
			}\\
			\subfloat[]{%
				\myincludegraphics[width=0.49\textwidth]{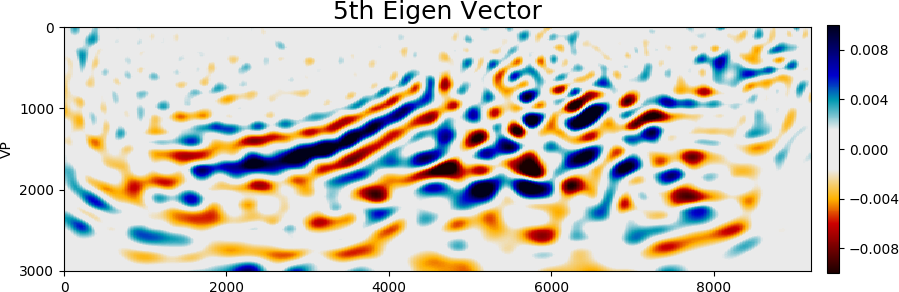}
				\myincludegraphics[width=0.49\textwidth]{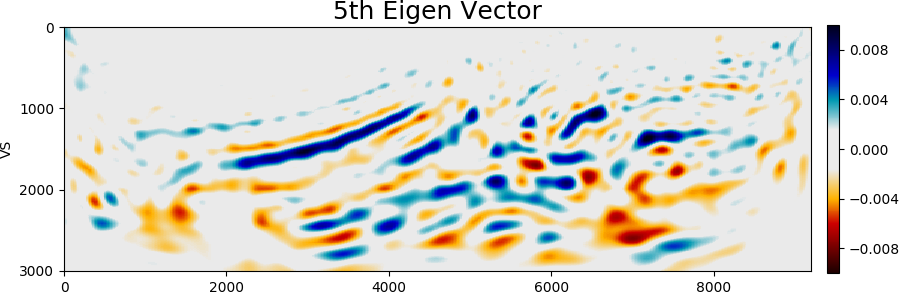}
			}\\
			\subfloat[]{%
				\myincludegraphics[width=0.49\textwidth]{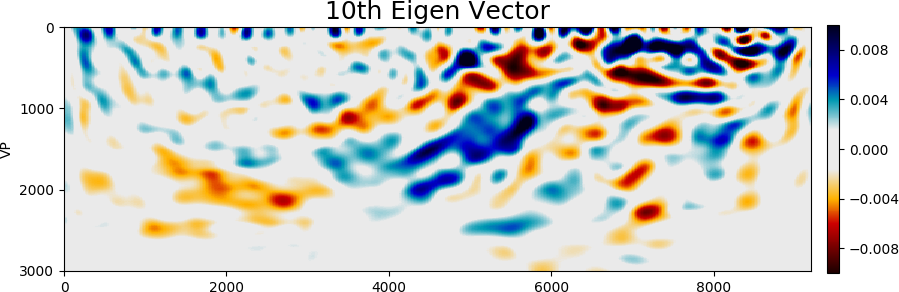}
				\myincludegraphics[width=0.49\textwidth]{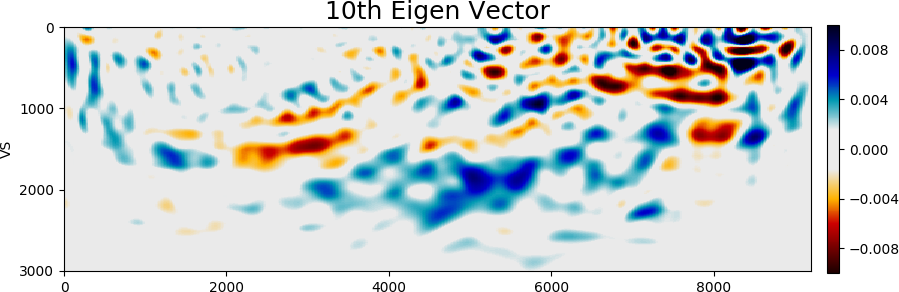}
			}\\
			\subfloat[]{%
				\myincludegraphics[width=0.49\textwidth]{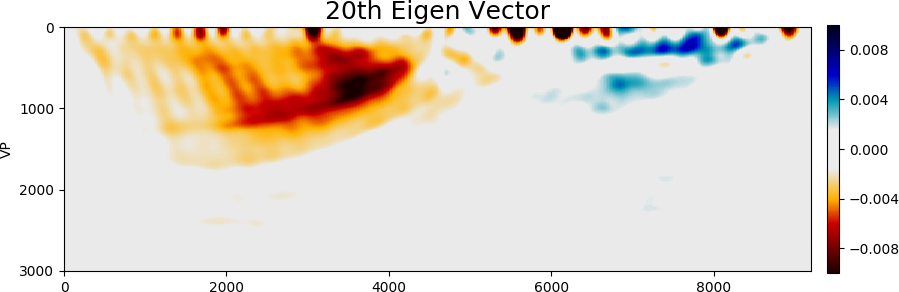}
				\myincludegraphics[width=0.49\textwidth]{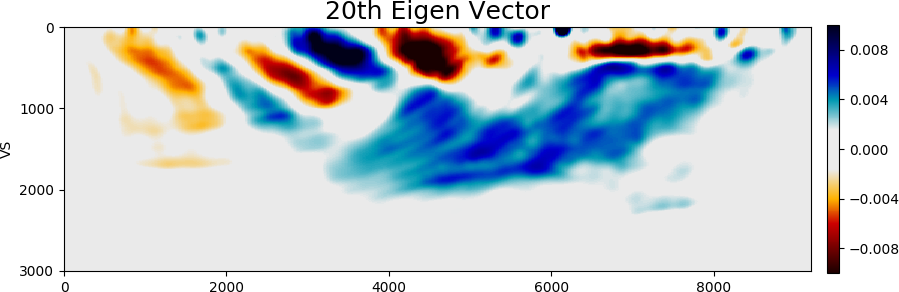}
			}\\
			\subfloat[]{%
				\myincludegraphics[width=0.49\textwidth]{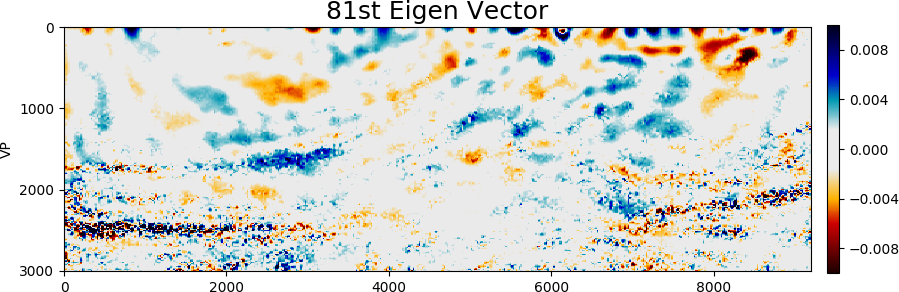}
				\myincludegraphics[width=0.49\textwidth]{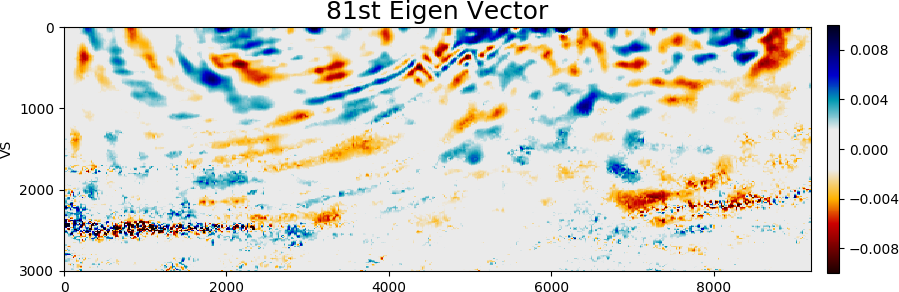}
			}
		\end{center}
		
		\caption{Eigendecomposition of the inverse Hessian from BFGS~\&~P2 elastic FWI using randomized SVD. Plotted in (a) are the eigenvalues on a logarithmic scale. Plotted in the remainder are the 1st, 5th, 10th, 20th, and the final eigenvectors, respectively. (Left column:~$V_P$\,; right column:~$V_S$\,.)}
	\end{figure}
	
	\begin{figure}
		\begin{center}
			\captionsetup[subfloat]{farskip=2pt,captionskip=2pt}
			\subfloat[]{%
				\myincludegraphics[width=0.42\textwidth]{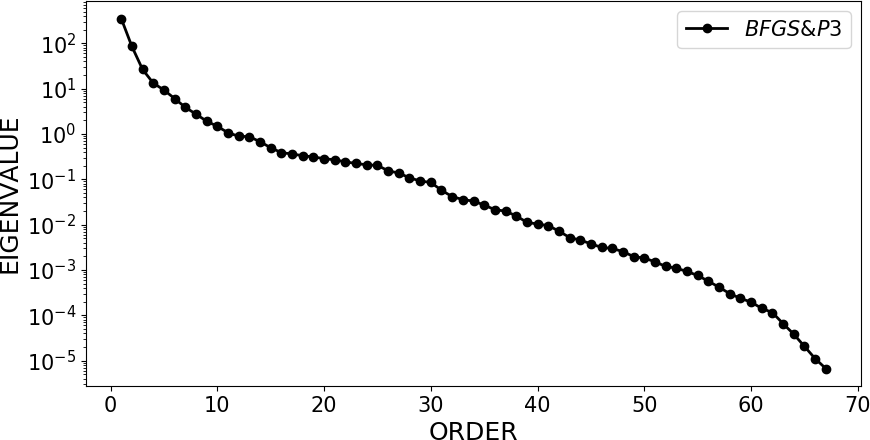}
			}\\
			\subfloat[]{%
				\myincludegraphics[width=0.49\textwidth]{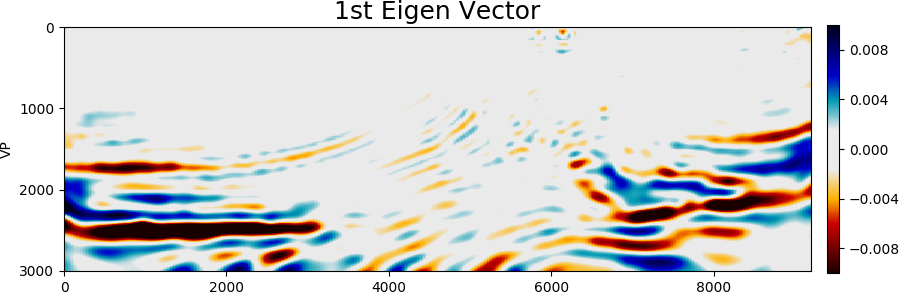}
				\myincludegraphics[width=0.49\textwidth]{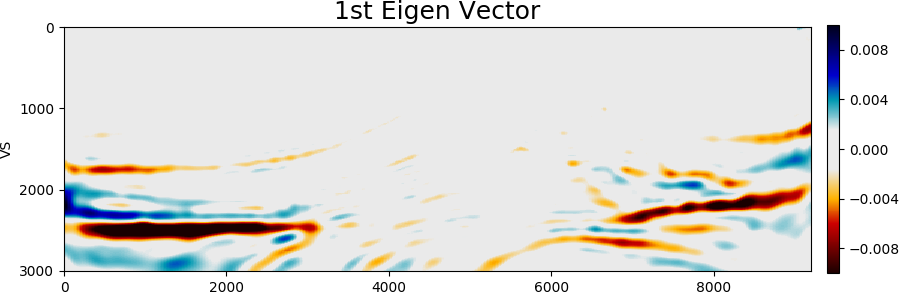}
			}\\
			\subfloat[]{%
				\myincludegraphics[width=0.49\textwidth]{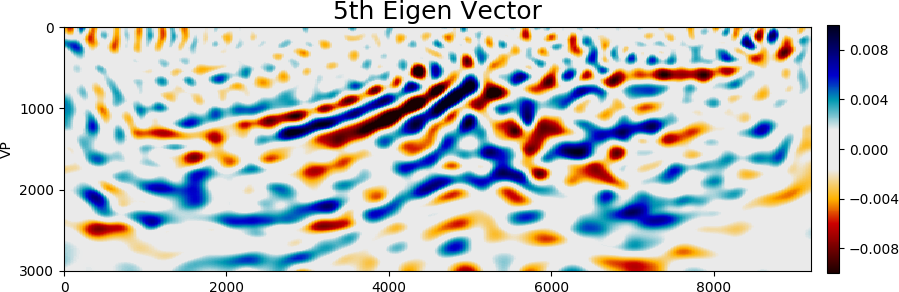}
				\myincludegraphics[width=0.49\textwidth]{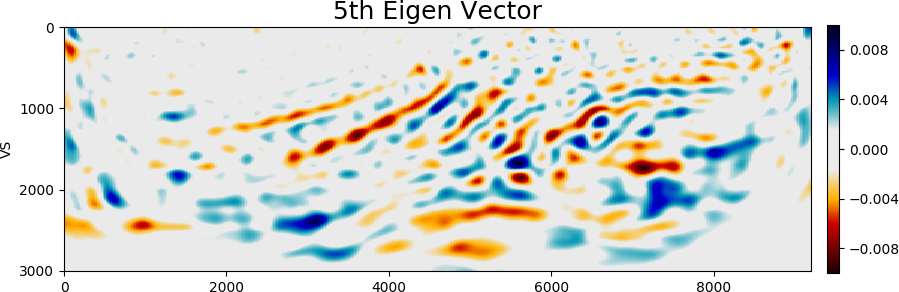}
			}\\
			\subfloat[]{%
				\myincludegraphics[width=0.49\textwidth]{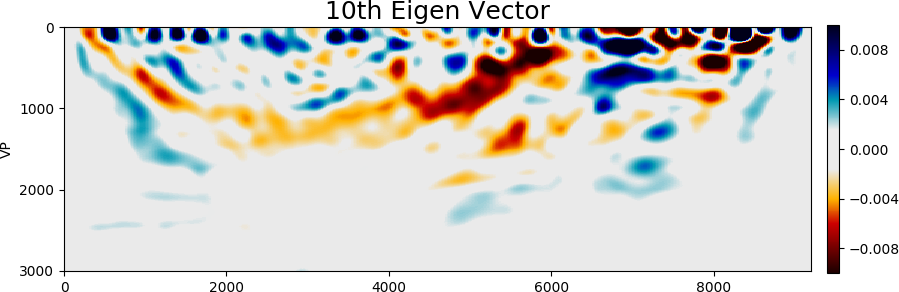}
				\myincludegraphics[width=0.49\textwidth]{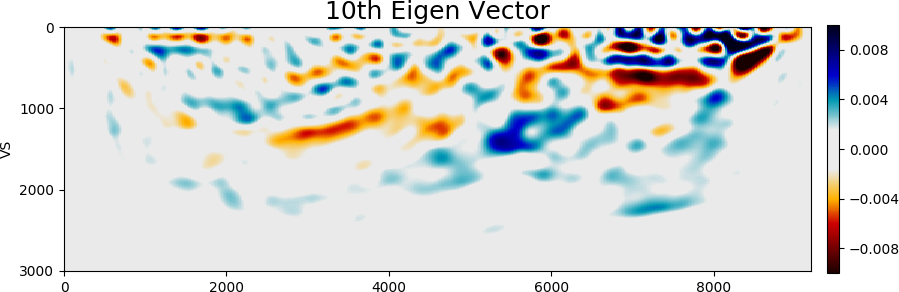}
			}\\
			\subfloat[]{%
				\myincludegraphics[width=0.49\textwidth]{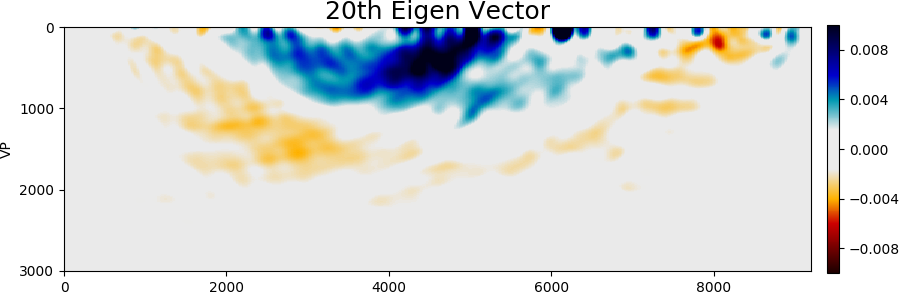}
				\myincludegraphics[width=0.49\textwidth]{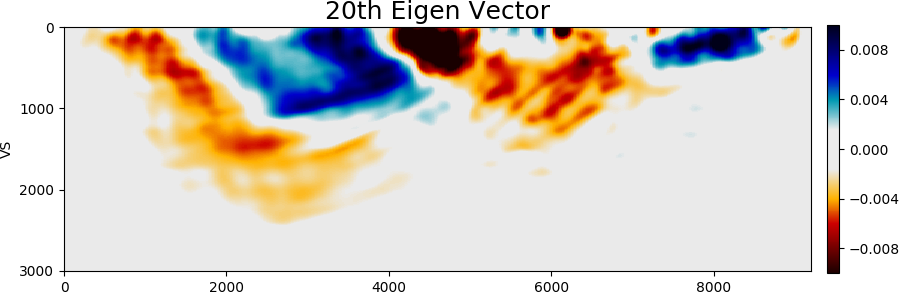}
			}\\
			\subfloat[]{%
				\myincludegraphics[width=0.49\textwidth]{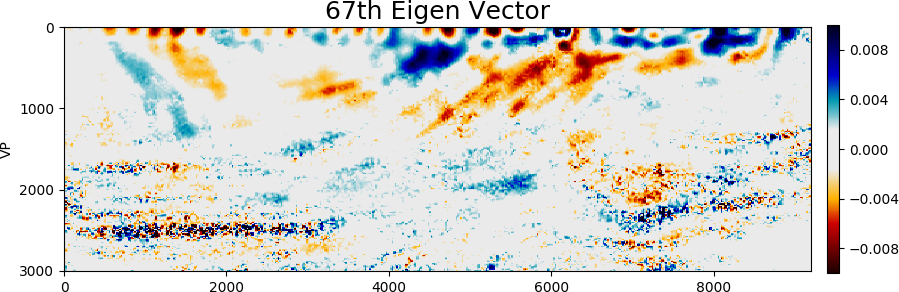}
				\myincludegraphics[width=0.49\textwidth]{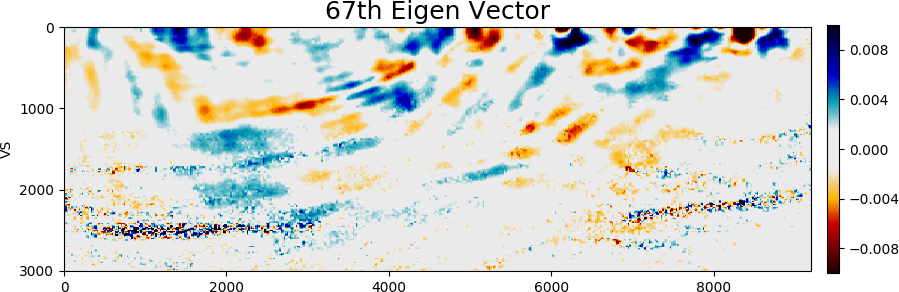}
			}
		\end{center}
		
		\caption{Eigendecomposition of the inverse Hessian from BFGS~\&~P3 elastic FWI using randomized SVD. Plotted in (a) are the eigenvalues on a logarithmic scale. Plotted in the remainder are the 1st, 5th, 10th, 20th, and the final eigenvectors, respectively. (Left column:~$V_P$\,; right column:~$V_S$\,.)}
	\end{figure}
	
	\begin{figure}
		\begin{center}
			\captionsetup[subfloat]{farskip=2pt,captionskip=2pt}
			\subfloat[]{%
				\myincludegraphics[width=0.42\textwidth]{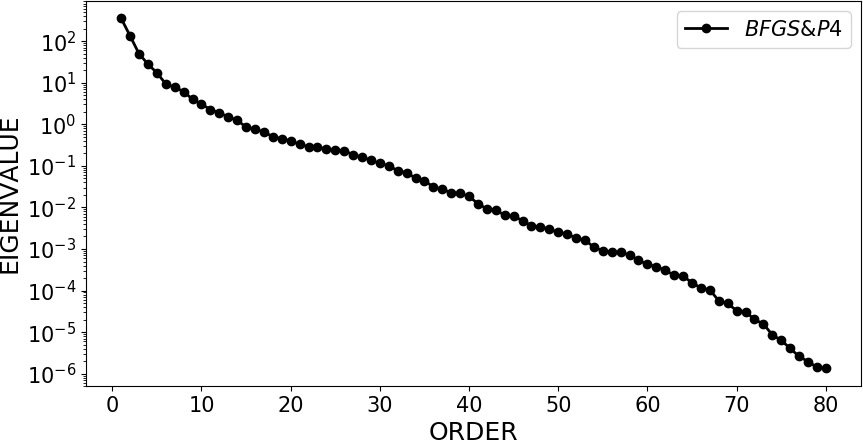}
			}\\
			\subfloat[]{%
				\myincludegraphics[width=0.49\textwidth]{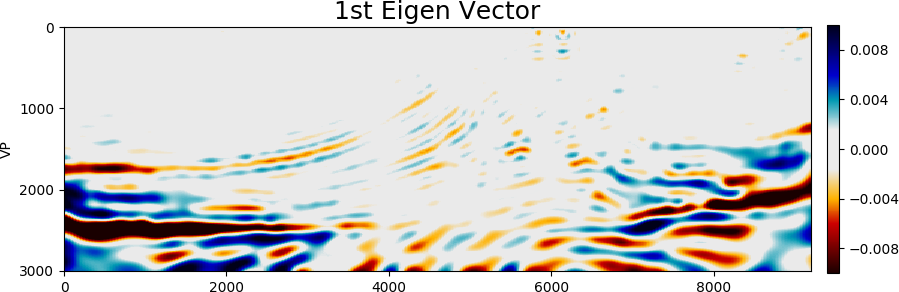}
				\myincludegraphics[width=0.49\textwidth]{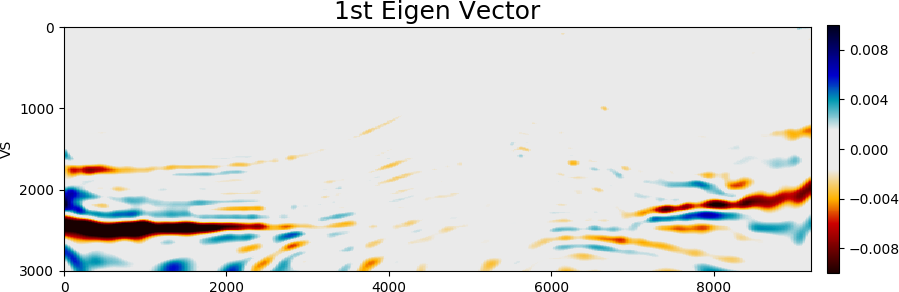}
			}\\
			\subfloat[]{%
				\myincludegraphics[width=0.49\textwidth]{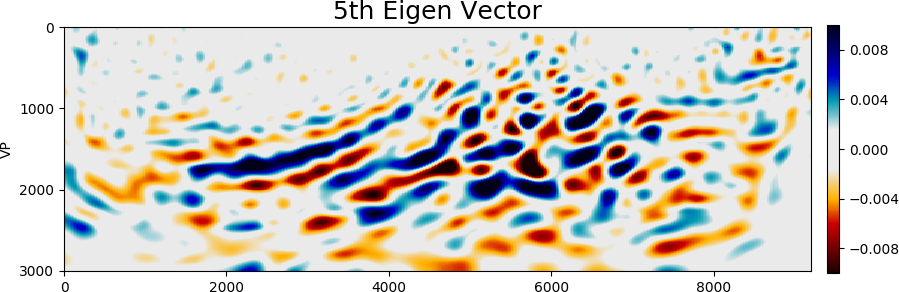}
				\myincludegraphics[width=0.49\textwidth]{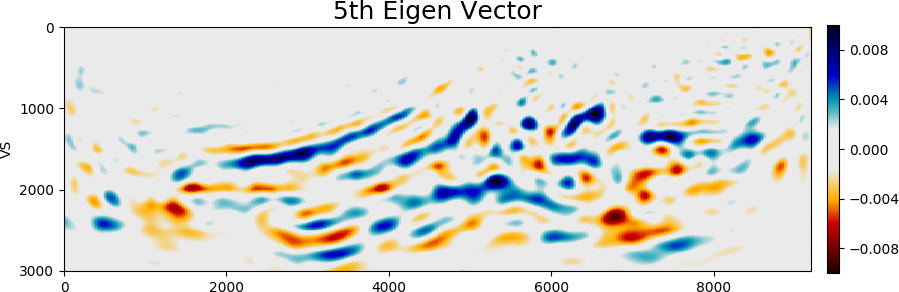}
			}\\
			\subfloat[]{%
				\myincludegraphics[width=0.49\textwidth]{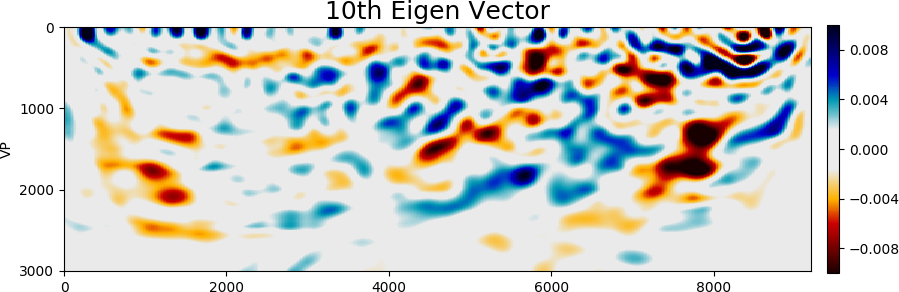}
				\myincludegraphics[width=0.49\textwidth]{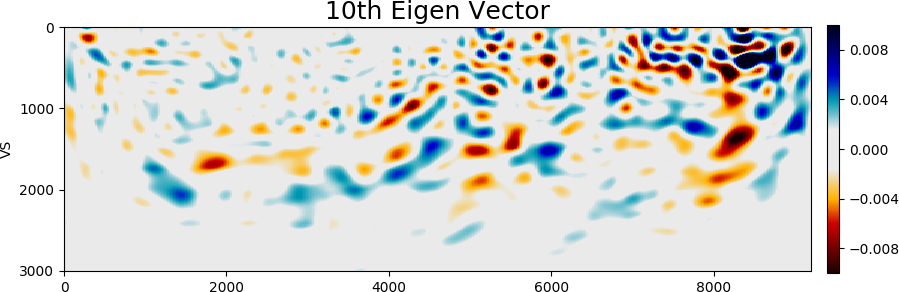}
			}\\
			\subfloat[]{%
				\myincludegraphics[width=0.49\textwidth]{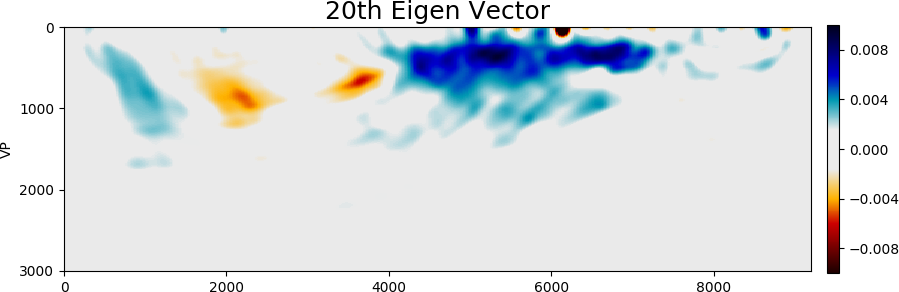}
				\myincludegraphics[width=0.49\textwidth]{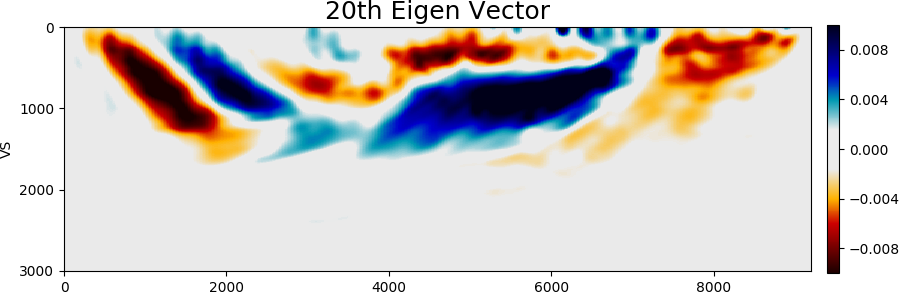}
			}\\
			\subfloat[]{%
				\myincludegraphics[width=0.49\textwidth]{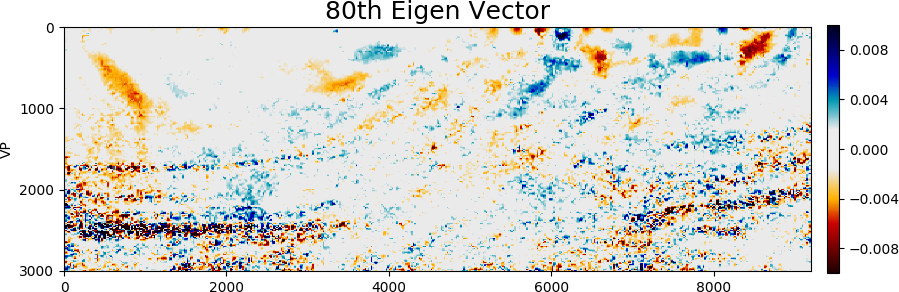}
				\myincludegraphics[width=0.49\textwidth]{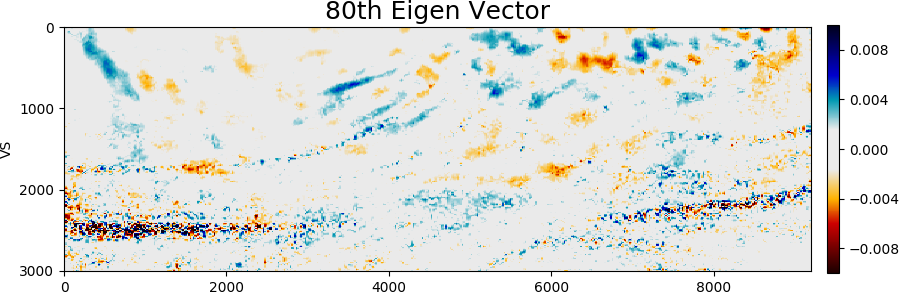}
			}
		\end{center}
		
		\caption{Eigendecomposition of the inverse Hessian from BFGS~\&~P4 elastic FWI using randomized SVD. Plotted in (a) are the eigenvalues on a logarithmic scale. Plotted in the remainder are the 1st, 5th, 10th, 20th, and the final eigenvectors, respectively. (Left column:~$V_P$\,; right column:~$V_S$\,.)}
	\end{figure}
	
	\begin{figure}
		\begin{center}
			\captionsetup[subfloat]{farskip=2pt,captionskip=2pt}
			\subfloat[]{%
				\myincludegraphics[width=0.49\textwidth]{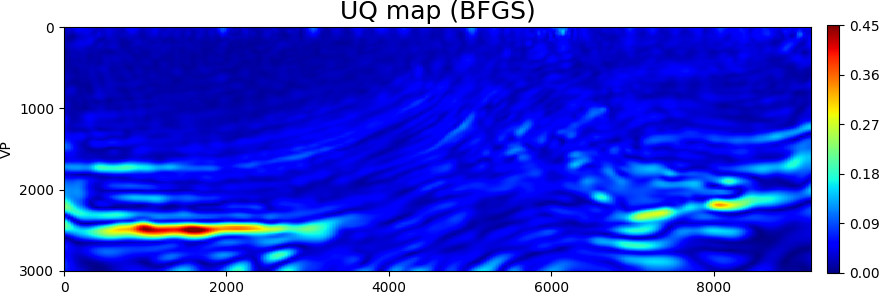}
				\myincludegraphics[width=0.49\textwidth]{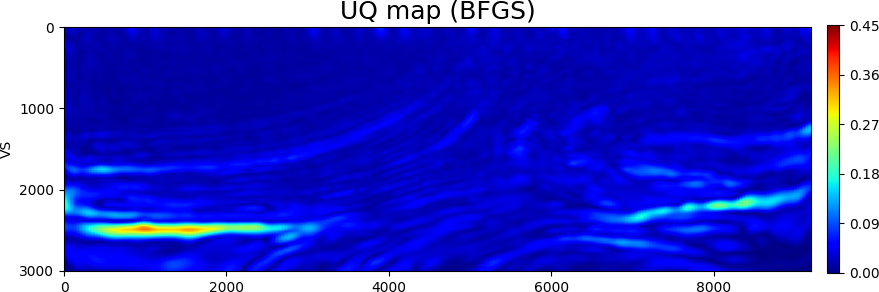}
			}\\
			\subfloat[]{%
				\myincludegraphics[width=0.49\textwidth]{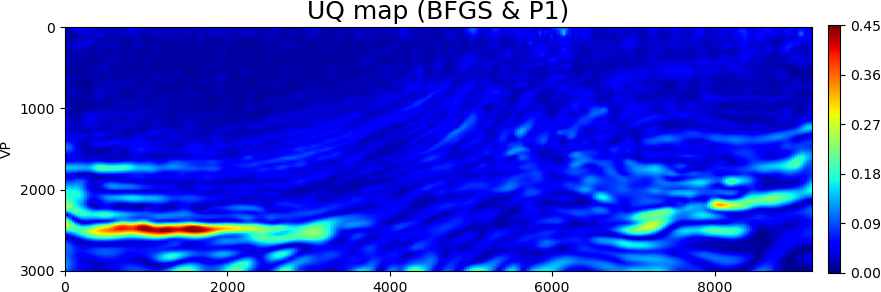}
				\myincludegraphics[width=0.49\textwidth]{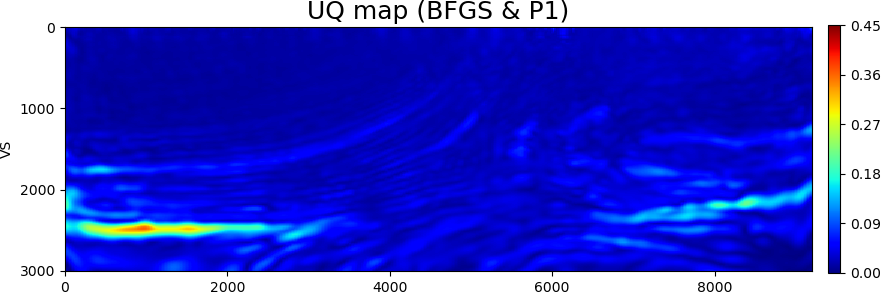}
			}\\
			\subfloat[]{%
				\myincludegraphics[width=0.49\textwidth]{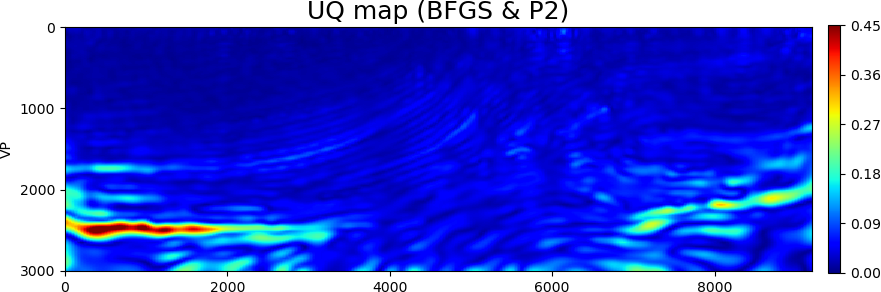}
				\myincludegraphics[width=0.49\textwidth]{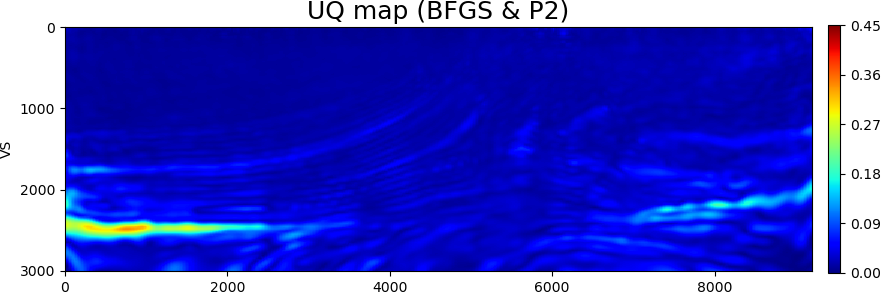}
			}\\
			\subfloat[]{%
				\myincludegraphics[width=0.49\textwidth]{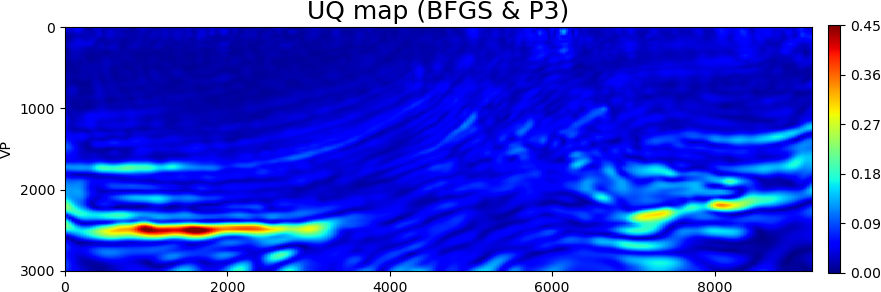}
				\myincludegraphics[width=0.49\textwidth]{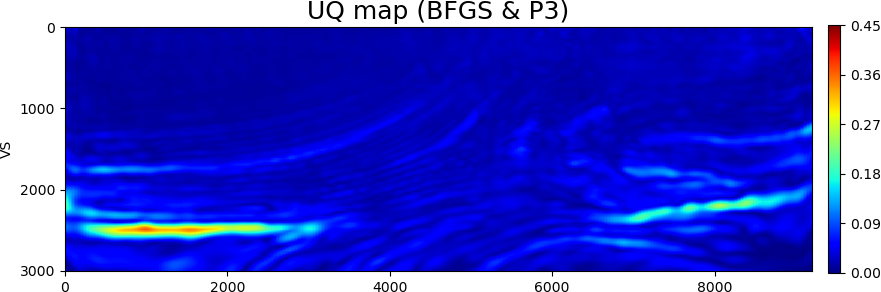}
			}\\
			\subfloat[]{%
				\myincludegraphics[width=0.49\textwidth]{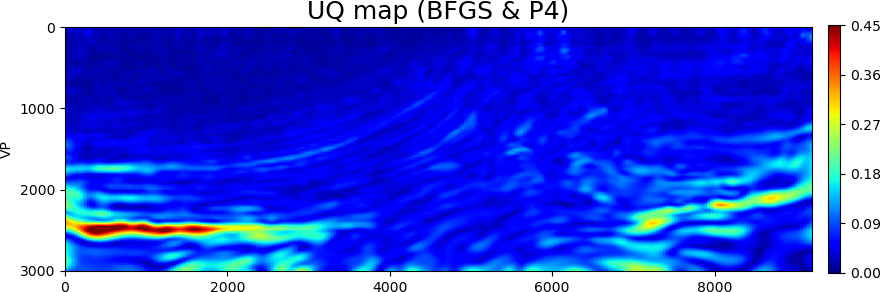}
				\myincludegraphics[width=0.49\textwidth]{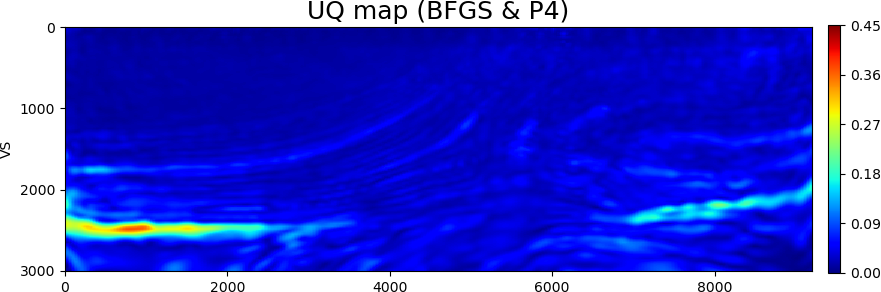}
			}
		\end{center}
		
		\caption{Uncertainty quantification maps for elastic FWI using non-preconditioned and preconditioned BFGS algorithms, as shown from top to bottom. Please see more details in Appendix C on how to estimate an absolute scaling range for the UQ maps. (Left column:~$V_P$\,; right column:~$V_S$\,.)}
	\end{figure}
	
	\begin{figure}
		\begin{center}
			\captionsetup[subfloat]{farskip=2pt,captionskip=2pt}
			{%
				\myincludegraphics[width=0.49\textwidth]{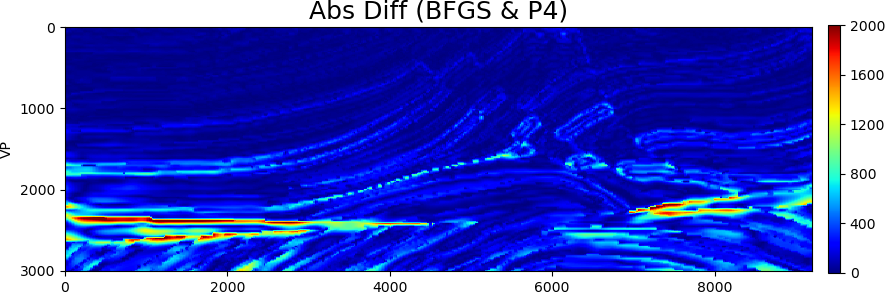}
				\myincludegraphics[width=0.49\textwidth]{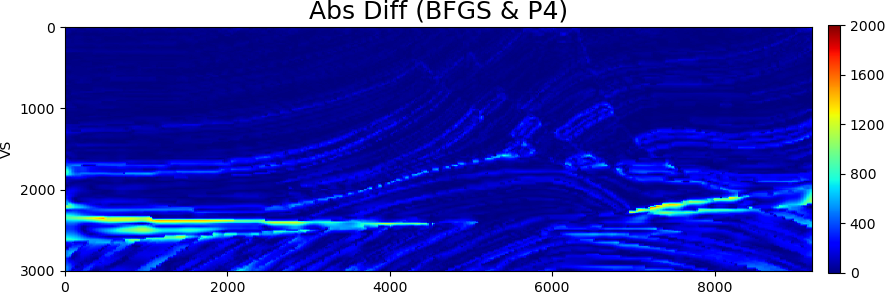}
			}
			
		\end{center}
		
		\caption{Absolute model difference between the true model in Fig.~1a and the inverted model in Fig.~4d for reference. The model differences shown here resemble the uncertainty quantification maps in Fig.~11. (Left column:~$V_P$\,; right column:~$V_S$\,.)}
	\end{figure}
	
	\clearpage

	\appendix
	
	\renewcommand\thesection{\Alph{section}}
	\renewcommand{\thefigure}{\thesection\arabic{figure}}
	
	\section{Determination of the diagonal Hessian}
	
	FWI aims to minimize the  misfit function
	\eq
	{f}({\bf{m}}) = \frac{1}{2}\left\|  \mathbf{s}(\mathbf{m}) -\bf{d}\right\|_2^2,
	\en
	in which~$\bf{d}$ are the observed data and~$\mathbf{s}(\mathbf{m})=\bf{Ru}(\bf{m})$ the simulated data, with~$\bf{u}(\bf{m})$ the simulated wavefield from model~${\bf{m}}$ and~${\bf{R}}$ the acquisition sampling operator. Using the Born approximation for a model perturbation~${\bf{m}} = {{\bf{m}}_0} + \delta {\bf{m}}$, we have
	\eq
	{\bf{s}}({{\bf{m}}_0} + \delta {\bf{m}}) = {\bf{s}}({{\bf{m}}_0}) + {\bf{J}}\delta {\bf{m}} + O(\delta {{\bf{m}}^2}),
	\en
	with Jacobian matrix~${\bf{J}} = {{\delta {\bf{s}}} \mathord{\left/{\vphantom {{\delta {\bf{d}}} {\delta {\bf{m}}}}} \right.\kern-\nulldelimiterspace} {\delta {\bf{m}}}}$.
	Taking the first-order derivative of eq.~(A1) yields
	\eq
	\frac{{\partial {{f}}}}{{\partial {\bf{m}}}} = {{\bf{J}}^\dag }\left[ {{\bf{J}}\left( {{{\bf{m}}_0} + \delta {\bf{m}}} \right) - {{\bf{d}}}} \right] = {{\bf{J}}^\dag }\left( {{\bf{J}}\delta {\bf{m}} - \delta {\bf{d}}} \right),
	\en
	with~$\delta {\bf{d}} = {{\bf{d}}} - {\bf{J}}{{\bf{m}}_0}$ and~$\dag$ the adjoint of an operator. When~${{\partial {{f}}} \mathord{\left/{\vphantom {{\partial {{f}}} {\partial {\bf{m}}}}} \right.\kern-\nulldelimiterspace} {\partial {\bf{m}}}} = 0$, we have a direct solution for model perturbation~$\delta {\bf{m}}$, namely,
	\eq
	\delta {\bf{m}} = {\left( {{{\bf{J}}^\dag }{\bf{J}}} \right)^{ - 1}}{{\bf{J}}^\dag }\delta {\bf{d}}.
	\en
	However, the Hessian~${\bf{H}} = {{\bf{J}}^\dag }{\bf{J}}$ is computationally too expensive to calculate, store, and invert. Hence, in practical applications, we simply calculate the model update as
	\eq
	\delta {{\bf{m}}^ * } = {{\bf{J}}^\dag }\delta {\bf{d}}.
	\en
	For convenience, let us use a velocity-stress formulation. The 2D velocity-stress isotropic elastic wave equation can be written as~\citep{vigh2014elastic,chen2017elastic}
	\eq
	\left( {\begin{array}{*{20}{c}}
			{\rho \,{\bf{I}}}&{\bf{0}}\\
			{\bf{0}}&{\bf{I}}
	\end{array}} \right)\,\partial_t \bf{u} - \left( {\begin{array}{*{20}{c}}
			{\bf{0}}&{{{\bf{D}}^T}}\\
			{{\bf{CD}}}&{\bf{0}}
	\end{array}} \right){\bf{u}} = {\bf{f}},
	\en
	where~${\bf{f}}$ is the source term, ${\bf{I}}$ the identity matrix, and
	\eq
	\begin{array}{l}
		{\bf{u}} = \left( {{\bf{v}},{\mathbf{\sigma }}} \right),\qquad {\bf{v}} = {({v_x},{v_z})^T},\qquad {\bf{\sigma }} = {({\sigma _{xx}},{\sigma _{xz}},{\sigma _{zz}})^T},\\
		{\bf{C}} = \left( {\begin{array}{*{20}{c}}
				{\lambda  + 2\mu }&\lambda &0\\
				\lambda &{\lambda  + 2\mu }&0\\
				0&0&\mu 
		\end{array}} \right),\qquad {\bf{D}} = \left( {\begin{array}{*{20}{c}}
				\partial_x & 0\\
				0 & \partial_z\\
				\partial_z & \partial_x
		\end{array}} \right)
	\end{array}
	\en
	with~${\bf{v}}$ containing the particle velocities, ${\bf{\sigma }}$ the stress elements, ${\bf{C}}$ the isotropic elastic tensor with~$\lambda$ and~$\mu$ being the Lam\'e parameters, ${\bf{D}}$ an operator of spatial derivatives, and initial condition of the form~${\bf{u}}\left( {{\bf{x}},t\;| \;t < 0} \right) = {\bf{0}}$. To avoid clutter, we drop the spatial and temporal dependence~${\bf u} = {\bf u}({\bf{x}},t)$ and the spatial dependence of~$\bf{C} = \bf{C}({\bf{x}})$. Eq.~(A7) is the state equation when FWI runs as an optimal control problem~\citep{Tromp2005,Plessix2006}. For simplification, the abstract form of eq.~(A7) reads
	\eq
	{\bf{Au}} = {\bf{f}},
	\en
	and its first-order derivative over~$\bf{m}$ reads
	\eq
	{\bf{A}}\,\frac{{\partial {\bf{u}}}}{{\partial {\bf{m}}}} =  \mbox{}- \frac{{\partial {\bf{A}}}}{{\partial {\bf{m}}}}\,{\bf{u}}.
	\en
	Now the Jacobian~${\bf{J}}$ in eq.~(A2) can be recast as
	\eq
	{\bf{J}} = \frac{{\partial {\bf{s}}}}{{\partial {\bf{m}}}} = {\bf{R}}\,\frac{{\partial {\bf{u}}}}{{\partial {\bf{m}}}} =  \mbox{}- {\bf{R}}{{\bf{A}}^{ - 1}}\,\frac{{\partial {\bf{A}}}}{{\partial {\bf{m}}}}\,{\bf{u}},
	\en
	in which~${{\bf{A}}^{ - 1}}$ is the Green's operator given that~${\bf{u}} = {{\bf{A}}^{ - 1}}{\bf{s}}$. The gradient from the data residual can be rewritten in detail~\citep{tarantola1988theoretical} as
	\eq
	\delta {{\bf{m}}^ * } = {{\bf{J}}^\dag }\delta {\bf{d}} = {\left( { \mbox{}- {\bf{R}}{{\bf{A}}^{ - 1}}\,\frac{{\partial {\bf{A}}}}{{\partial {\bf{m}}}}\,{\bf{u}}} \right)^\dag }\delta {\bf{d}} =  \mbox{}- {\left( {\frac{{\partial {\bf{A}}}}{{\partial {\bf{m}}}}\,{\bf{u}}} \right)^\dag }{\left( {{{\bf{A}}^\dag }} \right)^{ - 1}}\,{{\bf{R}}^\dag }\delta {\bf{d}},
	\en
	which represents the mapping from data residual~$\delta {\bf{d}}$ to the gradient~${\bf{J}}$ via the adjoint operator~${{\bf{J}}^\dag }$. By introducing the adjoint-state variable~$\widetilde {\bf{u}}$ from the following equation
	\eq
	{{\bf{A}}^\dag }\widetilde {\bf{u}} = {{\bf{R}}^\dag }\delta {\bf{d}},
	\en
	its corresponding elastic wave equation can be expressed as
	\eq
	\left( {\begin{array}{*{20}{c}}
			{\rho \,{\bf{I}}}&{\bf{0}}\\
			{\bf{0}}&{\bf{I}}
	\end{array}} \right){\left( \partial_t \right)^\dag }\widetilde {\bf{u}} - {\left( {\begin{array}{*{20}{c}}
				{\bf{0}}&{{{\bf{D}}^T}}\\
				{{\bf{CD}}}&{\bf{0}}
		\end{array}} \right)^\dag }\widetilde {\bf{u}} = {{\bf{R}}^\dag }\,\delta {\bf{d}},
	\en
	where~${{\bf{R}}^\dag }\delta {\bf{d}}$ acts as the adjoint source. After the operators of the adjoint~$\dag$ are applied, Eq.~(A13) can be recast as
	\eq
	\left( {\begin{array}{*{20}{c}}
			{\rho\, {\bf{I}}}&{\bf{0}}\\
			{\bf{0}}&{\bf{I}}
	\end{array}} \right)\left( \mbox{}- \partial_t\widetilde {\bf{u}}\right) + \left( {\begin{array}{*{20}{c}}
			{\bf{0}}&{{{\bf{D}}^T}{\bf{C}}}\\
			{\bf{D}}&{\bf{0}}
	\end{array}} \right)\widetilde {\bf{u}} = {{\bf{R}}^\dag }\,\delta {\bf{d}},
	\en
	The term~$\delta {\bf{\sigma }}$ in~$\delta {\bf{d}} = \left( {\delta {\bf{v}},\delta {\bf{\sigma }}} \right)$ remains zero because we only observe~$\delta {\bf{v}}$ in practice. The final form of the adjoint-state equation~\citep{vigh2014elastic} is
	\eq
	\left( {\begin{array}{*{20}{c}}
			{\rho \,{\bf{I}}}&{\bf{0}}\\
			{\bf{0}}&{\bf{I}}
	\end{array}} \right)\left( \mbox{}-\partial_t \widetilde {\bf{u}}\right) + \left( {\begin{array}{*{20}{c}}
			{\bf{0}}&{{{\bf{D}}^T}}\\
			{{\bf{CD}}}&{\bf{0}}
	\end{array}} \right)\widetilde {\bf{u}} = {{\bf{R}}^\dag }\left( \begin{array}{c}
		\delta {\bf{v}}\\
		{\bf{0}}
	\end{array} \right).
	\en
	Let us go back to the gradient in eq.~(A5), which can be further expressed as
	\eq
	\delta {{\bf{m}}^ * } =  \mbox{}- {\left( {\frac{{\partial {\bf{A}}}}{{\partial {\bf{m}}}}\,{\bf{u}}} \right)^\dag }\widetilde {\bf{u}},
	\en
	or in an explicit form as
	\eq
	\delta {\bf{m}}{\left( {\bf{x}} \right)^ * } =  \mbox{}- \int {{{\left( {\frac{{\partial {\bf{A}}}}{{\partial {\bf{m}}}}\,{\bf{u}}} \right)}^\dag }} \widetilde {\bf{u}} \,\mathrm{d}t =  \mbox{}- \int {\left( {\frac{{\partial {\bf{A}}}}{{\partial {\bf{m}}}}\,{\bf{u}}} \right)} \widetilde {\bf{u}}{^T}\,\mathrm{d}t,
	\en
	with~$\widetilde {\bf{u}}{\left( {{\bf{x}},t} \right)^T} = \widetilde {\bf{u}}{\left( {{\bf{x}},t} \right)^\dag }$ applied.  Eq.~(A17) is well known in reverse-time migration~\citep{baysal1983reverse} with the adjoint method. 
	
	When computing the gradients, we may have different parametrizations, such as~$\left( {\alpha ,\beta ,\rho } \right)$ or~$\left( {\lambda ,\mu ,\rho } \right)$, with~$\alpha ,\beta ,\rho~$ being the P- and S-wave wavespeeds, and densities, $\lambda ,\mu$ being Lam\'e parameters, respectively. We express the gradients with respect to~$(\alpha ,\beta)$ as follows
	\eq
	\begin{array}{l}
		\delta {\alpha ^*} = \int {\left( {\frac{{\partial {\bf{C}}}}{{\partial \alpha }}\,{\bf{Dv}}} \right) \cdot \widetilde {\bf{\sigma }}\,\mathrm{d}t} ,\\
		\delta {\beta ^*} = \int {\left( {\frac{{\partial {\bf{C}}}}{{\partial \beta }}\,{\bf{Dv}}} \right) \cdot \widetilde {\bf{\sigma }}\,\mathrm{d}t} .\\
	\end{array}
	\en
	
	Considering eq.~(A4), we aim to accelerate FWI with the Hessian in a Gauss-Newton approximation, which has the form
	\eq
	{\bf{H}} = {{\bf{J}}^\dag }{\bf{J}} = {\left( {{\bf{R}}{{\bf{A}}^{ - 1}}\,\frac{{\partial {\bf{A}}}}{{\partial {\bf{m}}}}\,{\bf{u}}} \right)^\dag }\left( {{\bf{R}}{{\bf{A}}^{ - 1}}\,\frac{{\partial {\bf{A}}}}{{\partial {\bf{m}}}}\,{\bf{u}}} \right),
	\en
	in which~${{\bf{A}}^{ - 1}}$ resembles the Green's operator and~${\bf{u}}$ the source wavefield. Regardless of the band-limited source wavelet, ${\bf{u}}$~resembles the source-side Green's functions~${{\bf{G}}_S}$, and at the same time, ${\bf{R}}{{\bf{A}}^{ - 1}}$ the receiver-side Green's functions~${{\bf{G}}_R}$.
	
	The full computation and storage of~${\bf{H}}$ are prohibitive. Therefore, \cite{shin2008improved}~propose to forget~${{{\bf{G}}_R}}$ to save computational cost, and only compute the Hessian diagonals via zero-lag cross-correlation of the source wavefields as follows:
	\eq
	{\rm H}({\bf{x}},{\bf{x}}) = \int {\left[ {\frac{{\partial {\bf{C}}}}{{\partial {m_i}}}\,{\bf{Dv}}\left( {\bf{x}} \right)} \right] \cdot \left[ {\frac{{\partial {\bf{C}}}}{{\partial {m_i}}}\,{\bf{Dv}}\left( {\bf{x}} \right)} \right]\,\mathrm{d}t}.
	\en
	When it comes to the parametrization~$\left( {\alpha ,\beta} \right)$, we have
	\eq
	\begin{array}{l}
		{{\rm H}_{\alpha \alpha }} = \int {\left( {\frac{{\partial {\bf{C}}}}{{\partial \alpha }}\,{\bf{Dv}}} \right) \cdot \left( {\frac{{\partial {\bf{C}}}}{{\partial \alpha }}\,{\bf{Dv}}} \right)dt = 8{\rho ^2}{\alpha ^2}\int {{{\left( \partial_x v_x + \partial_z v_z \right)}^2}\,\mathrm{d}t} } ,\\
		{{\rm H}_{\beta \beta }} = \int {\left( {\frac{{\partial {\bf{C}}}}{{\partial \beta }}\,{\bf{Dv}}} \right) \cdot \left( {\frac{{\partial {\bf{C}}}}{{\partial \beta }}\,{\bf{Dv}}} \right)\,\mathrm{d}t}  = 16{\rho ^2}{\beta ^2}\int {\left[ {{{\left( \partial_x v_x \right)}^2} + {{\left( \partial_z v_z \right)}^2}} \right]dt + 4{\rho ^2}{\beta ^2}\int {{{\left( \partial_x v_z + \partial_z v_x \right)}^2}\,\mathrm{d}t} } ,\\
		{{\rm H}_{\rho \rho }} = \int {\left( {\partial_t\bf{v}} \right) \cdot \left( {\partial_t\bf{v}} \right)\,\mathrm{d}t}  = \int {\left[ {{{\left( \partial_t v_x \right)}^2} + {{\left( \partial_t v_z \right)}^2}} \right]\,\mathrm{d}t} ,
	\end{array}
	\en
	with~${\bf{v}}$ being the source wavefield in particle velocity. The third equation in Eq.~(A21) corresponds to a generalized ``ray density'' Hessian kernel~\citep{Luo2012} regarding ~$\rho$ in the parametrization ~$\left( {\lambda ,\mu, \rho} \right)$. Eq.~(A21), however, does not involve the receiver-side Green's functions. Following~\cite{Luo2012} and~\cite{Modrak2016}, we modify eq.~(A21) to include~${{\bf{G}}_R}$ as
	\eq
	\begin{array}{l}
		{\bar {\rm H}_{\alpha \alpha }} = \left| {\int {\left( {\frac{{\partial {\bf{C}}}}{{\partial \alpha }}\,{\bf{Dv}}} \right) \cdot \left( {\frac{{\partial {\bf{C}}}}{{\partial \alpha }}\,{\bf{D}}\bar {\bf{v}}} \right)\,\mathrm{d}t} } \right| = 8{\rho ^2}{\alpha ^2}\left| {\int {\left( \partial_x v_x + \partial_z v_z \right)\left( \partial_x {\bar v}_x + \partial_z {\bar v}_z \right)\,\mathrm{d}t} } \right|,\\
		{\bar {\rm H}_{\beta \beta }} = \left| {\int {\left( {\frac{{\partial {\bf{C}}}}{{\partial \beta }}\,{\bf{Dv}}} \right) \cdot \left( {\frac{{\partial {\bf{C}}}}{{\partial \beta }}\,{\bf{D}}\bar {\bf{v}}} \right)\,\mathrm{d}t} } \right| \\
		\; \; \; \; \; \; \,= 16{\rho ^2}{\beta ^2}\left| {\int {\left( \partial_x v_x\,\partial_x {\bar v}_x + \partial_z v_z\,\partial_z {\bar v}_z \right)\,\mathrm{d}t} } \right| + 4{\rho ^2}{\beta ^2}\left| {\int {\left( \partial_x v_z + \partial_z v_x \right)\left( \partial_x {\bar v}_z + \partial_z {\bar v}_x \right)\,\mathrm{d}t} } \right|,\\
		{\bar {\rm H}_{\rho \rho }} = \left| {\int {\left( {\partial_t\bf{v}} \right) \cdot \left( \partial_t \bar{\bf{v}} \right)\,\mathrm{d}t} } \right|  =  \left| {\int {\left( \partial_t v_x\,\partial_t {\bar v}_x + \partial_t v_z\,\partial_t {\bar v}_z \right)\,\mathrm{d}t} } \right|,
	\end{array}
	\en
	with~$\bar {\bf{v}}$ denoting the receiver wavefield in particle velocity. Note that although both wavefields originate at the receivers, $\bar {\bf{v}}$ differs from~$\widetilde {\bf{v}}$ in that the former indicates the wavefield due to the simulated data while the latter the wavefield due to the data residual. We take their absolute values to ensure positive definiteness of the initial Hessian.
	
	\section{Validation with the elastic Overthrust model}
	
	This section is to validate the aforementioned preconditioned-BFGS based elastic FWI and uncertainty quantification workflow using the elastic Overthrust model, which can be found in \cite{liu2019a}. The simulations are with sources and receivers 10-m deep and using absorbing boundaries \citep{stacey1988improved}. For simplicity, we only use preconditioner P4. Shown in Fig.~B1 are the true, initial, and inverted elastic Overthrust models. Also shown are the associated absolute model differences between the true and inverted model together with the corresponding uncertainty maps. We observe that Fig.~B1d resembles Fig.~B1e.
	
	\begin{figure}
		\begin{center}
			\captionsetup[subfloat]{farskip=2pt,captionskip=2pt}
			\subfloat[]{%
				\myincludegraphics[width=0.49\textwidth]{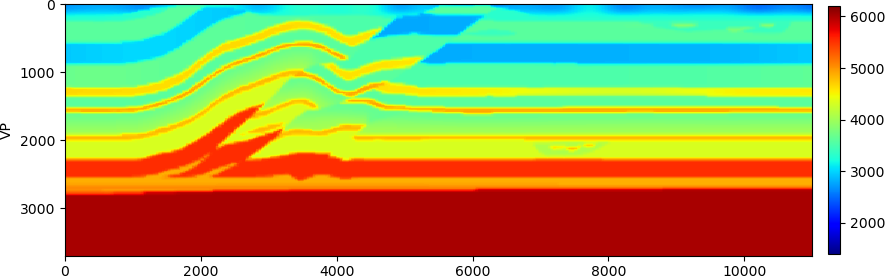}
				\myincludegraphics[width=0.49\textwidth]{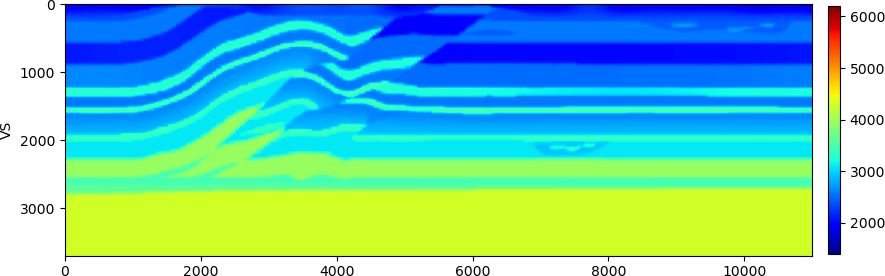}
			}\\
			\subfloat[]{%
				\myincludegraphics[width=0.49\textwidth]{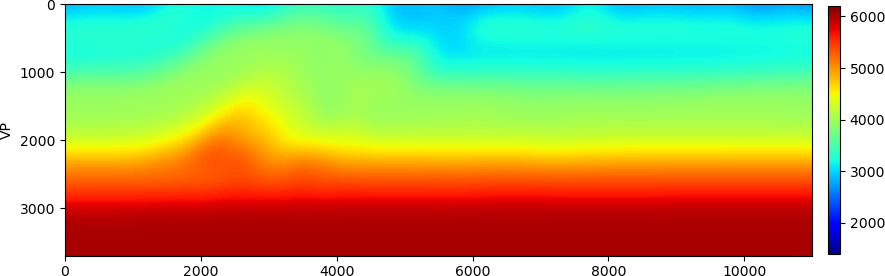}
				\myincludegraphics[width=0.49\textwidth]{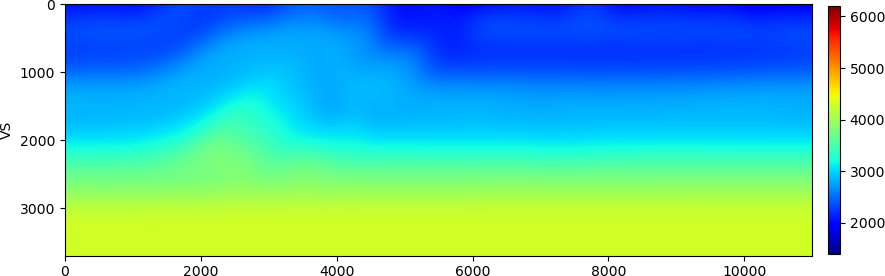}
			}\\
			\subfloat[]{%
				\myincludegraphics[width=0.49\textwidth]{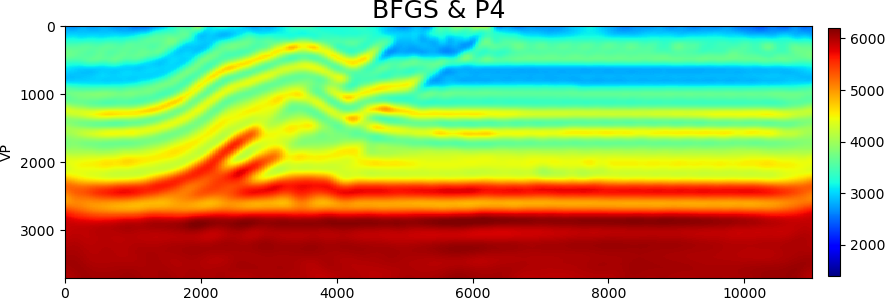}
				\myincludegraphics[width=0.49\textwidth]{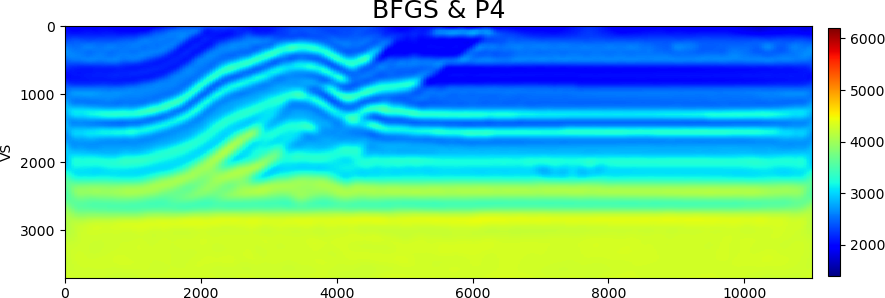}
			}\\
			\subfloat[]{%
				\myincludegraphics[width=0.49\textwidth]{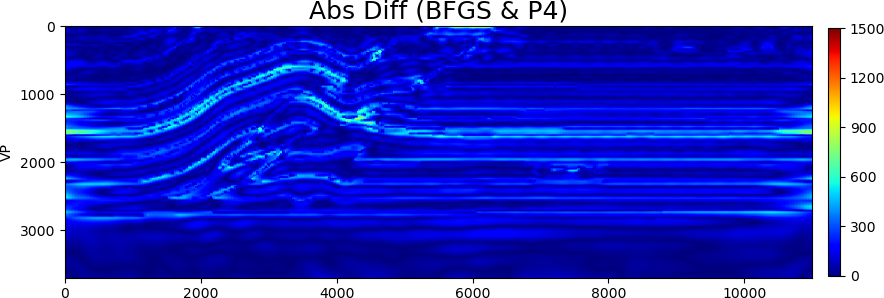}
				\myincludegraphics[width=0.49\textwidth]{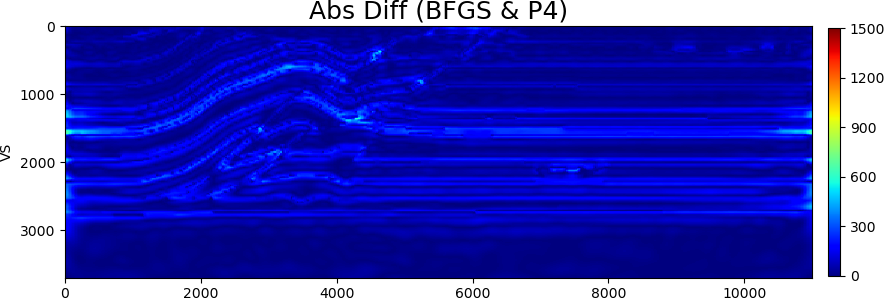}
			}\\
			\subfloat[]{%
				\myincludegraphics[width=0.49\textwidth]{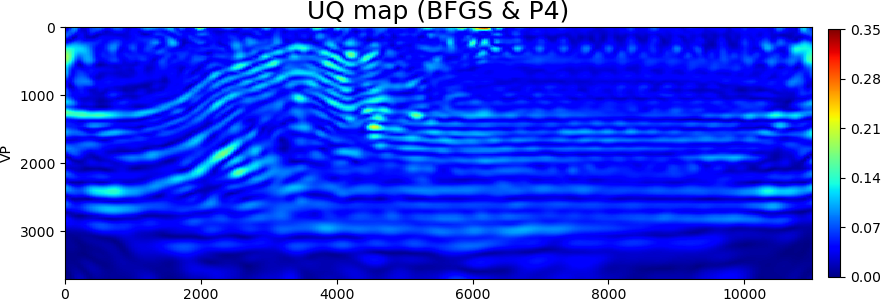}\hspace{1mm}
				\myincludegraphics[width=0.49\textwidth]{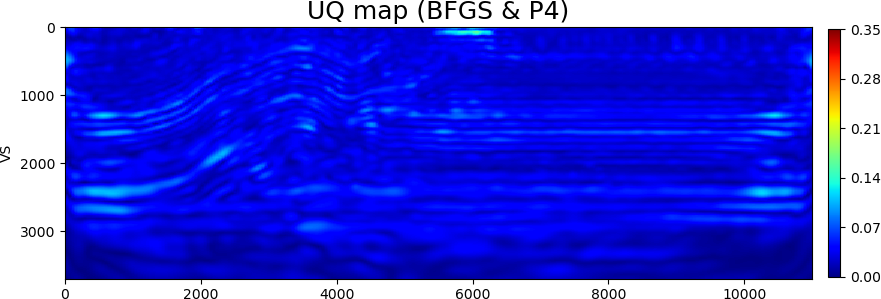}
			}
		\end{center}
		
		\caption{From top to bottom: (a) true and (b) initial and (c) inverted elastic Overthrust models, (d) the absolute model differences between the true and the inverted, (e) the uncertainty maps estimated from the preconditioned-BFGS FWI workflow with P4. (Left column:~$V_P$\,; right column:~$V_S$\,.)}
	\end{figure}
	\clearpage 
	\newpage
	\section{The scaling of the Uncertainty maps}
	The scaling of the UQ maps in Fig.~11, which mainly reflects the square-root diagonals of the ${\bf{H}}^{-1}$ term in Eqs.~12 and 13, ranges between [0,1] because all the gradients have been rescaled with a constant factor to a physically reasonable range. This factor comes from a separate trial once and for all per model, in which we determine the constant by performing a parabolic line search with the 1st unpreconditioned, smoothed gradient. We can, however, choose to scale for the UQ maps like the way in the SRVM-based null-space shuttle. \citep{liu2020square}. The data misfits with respect to the UQ-map-based perturbations are shown in Fig.~C1, in which one $m_{pert}$ is scaled from in Fig.~11e to be with a maximum of 250~$m/s$. 
	
	Another straightforward way is to get the UQ map scaling by filtering ${{\bf{H}}^{ - 1}}$  through the model prior covariance. As indicated by Eqs.~12 and 13, when there is no data information gain, ${\bf{H}}^{-1} \rightarrow \bf{I}$; otherwise, the scaled UQ map will look like the ones in Fig.~11 but with a scaling from the model prior covariance. \\
	\setcounter{figure}{0}
	\begin{figure}
	\begin{center}
	\myincludegraphics[width=0.59\textwidth]{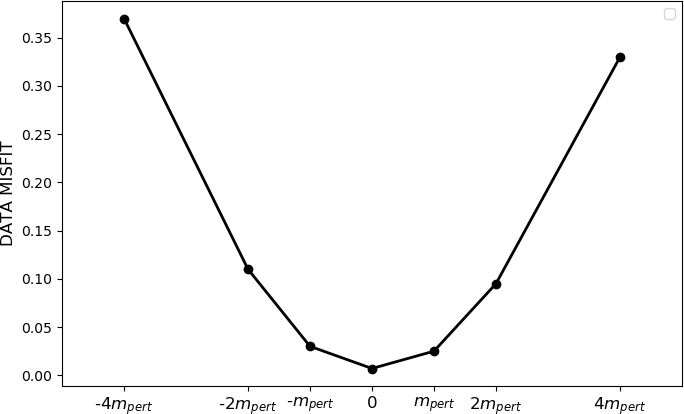}
	\end{center}
	
	\caption{Data misfits with respect to the UQ-map-based model perturbations. The UQ map here refers to that in Fig.~11e, from which $m_{pert}$ is scaled to have a maximum of 250~$m/s$. The fact that the data-misfit curve remains flat within the range of [$-m_{pert}$, $m_{pert}$] indicates that we can roughly estimate the scaling of the UQ map using the posterior sampling methods stated in Appendix C. }
\end{figure}

\end{document}